\documentclass[aps,prx,twocolumn,10pt,superscriptaddress,nofootinbib,balancelastpage]{revtex4-1} 
\usepackage[latin1]{inputenc}
\usepackage{amsmath,amssymb}
\usepackage{mathrsfs} 
\usepackage[capitalise]{cleveref}
\usepackage{siunitx}
\usepackage{braket}
\usepackage{color}

\allowdisplaybreaks

\newcommand{\id}{\operatorname{id}}%
\newcommand{\ii}{\mathrm{i}}%
\newcommand{\dif}{\mathrm{d}}%
\newcommand{\tdif}[2]{\frac{\dif#1}{\dif#2}}%
\newcommand{\pdif}[2]{\frac{\partial#1}{\partial#2}}%
\newcommand{\Rea}{\operatorname{Re}}%
\newcommand{\Ima}{\operatorname{Im}}%
\newcommand{\norm}[1]{\lVert#1\rVert}%
\newcommand{\Tr}{\operatorname{Tr}}%
\newcommand{\ZT}[1]{\textquotedblleft#1\textquotedblright}%
\newcommand{\AS}{A_{\mathrm{S}}}%
\newcommand{\ASf}[1]{A_{\mathrm{S},#1}}%
\newcommand{\HH}{H_{\mathrm{H}}}%
\newcommand{\HS}{H_{\mathrm{S}}}%
\newcommand{\HSO}{H_{\mathrm{S},0}}%
\newcommand{\HL}{H_{\mathrm{l}}}%
\newcommand{\HSL}{H_{\mathrm{sl}}}%
\newcommand{\LH}{L_{\mathrm{H}}}%
\newcommand{\LS}{L_{\mathrm{S}}}%
\newcommand{\LSO}{L_{\mathrm{S},0}}%
\newcommand{\LSI}{L_{\mathrm{S},1}}%
\newcommand{\LLI}{L_{L,I}}%
\newcommand{\LRI}{L_{R,I}}%
\newcommand{\Ef}[1]{E_{#1}}%
\newcommand{\WW}{W}%
\newcommand{\ww}{w}%

\newlength{\myl}%
\newcommand{\INT}[3]{\settowidth{\myl}{$\displaystyle\int_{#1}^{#2}$}{\int_{#1}^{#2}\;\;\;\hspace{-\the\myl}\dif #3}\,}
\newcommand{\TINT}[3]{\settowidth{\myl}{$\int_{#1}^{#2}$}{\int_{#1}^{#2}\;\;\;\hspace{-\the\myl}\dif #3}\,}
\newcommand{\EINT}[3]{\settowidth{\myl}{$\int_{#1}^{#2}$}{\int_{#1}^{#2}\;\;\;\,\hspace{-\the\myl}\dif #3}\,}

\begin{document}

\title{Mori-Zwanzig projection operator formalism for systems with time-dependent Hamiltonians}
\author{Michael te Vrugt}
\affiliation{Institut f\"ur Theoretische Physik, Westf\"alische Wilhelms-Universit\"at M\"unster, D-48149 M\"unster, Germany}
\affiliation{Center for Soft Nanoscience, Westf\"alische Wilhelms-Universit\"at M\"unster, D-48149 M\"unster, Germany}

\author{Raphael Wittkowski}
\email[Corresponding author: ]{raphael.wittkowski@uni-muenster.de}
\affiliation{Institut f\"ur Theoretische Physik, Westf\"alische Wilhelms-Universit\"at M\"unster, D-48149 M\"unster, Germany}
\affiliation{Center for Soft Nanoscience, Westf\"alische Wilhelms-Universit\"at M\"unster, D-48149 M\"unster, Germany}
\affiliation{Center for Nonlinear Science, Westf\"alische Wilhelms-Universit\"at M\"unster, D-48149 M\"unster, Germany}

\begin{abstract}
The Mori-Zwanzig projection operator formalism is a powerful method for the derivation of mesoscopic and macroscopic theories based on known microscopic equations of motion. It has applications in a large number of areas including fluid mechanics, solid-state theory, spin relaxation theory, and particle physics. In its present form, however, the formalism cannot be directly applied to systems with time-dependent Hamiltonians. Such systems are relevant in a lot of scenarios like, for example, driven soft matter or nuclear magnetic resonance. In this article, we derive a generalization of the present Mori-Zwanzig formalism that is able to treat also time-dependent Hamiltonians. The extended formalism can be applied to classical and quantum systems, close to and far from thermodynamic equilibrium, and even in the case of explicitly time-dependent observables. Moreover, we develop a variety of approximation techniques that enhance the practical applicability of our formalism. Generalizations and approximations are developed for both equations of motion and correlation functions. Our formalism is demonstrated for the important case of spin relaxation in a time-dependent external magnetic field. The Bloch equations are derived together with microscopic expressions for the relaxation times.
\end{abstract}

\maketitle

\section{Introduction}
The Mori-Zwanzig projection operator formalism \cite{Mori1965, Zwanzig1960, Grabert1982, Zwanzig2001, Forster1989, HansenMD2009} is a central tool of statistical physics. It is based on the observation that macroscopic systems are typically well described by a small number of \textit{relevant} variables, even though they have a large number of microscopic degrees of freedom \cite{Grabert1978}. Typical examples include fluids, where the relevant variables are mass, momentum, and energy densities \cite{Forster1989, Sasa2014}, or spin systems, where the relevant variable is the magnetization \cite{KivelsonO1974}. The key idea is to introduce a projection operator that projects the full microscopic dynamics of the system onto the subspace that depends only on the relevant variables. Thereby, one obtains closed equations of motion for the relevant variables, in which the irrelevant dynamics appears as a noise term \cite{Grabert1982, Givon2005}.

This allows to derive mesoscopic and macroscopic theories based on known microscopic equations of motion in a systematic and rather compact way of coarse graining \cite{HijonEVEDB2010, Espanol2004, EspanolD2015}. Therefore, the Mori-Zwanzig formalism is a very useful method for a variety of fields, such as fluid mechanics \cite{Grabert1982, Forster1989}, polymer physics \cite{LiLDK2017, HijonEVEDB2010}, classical dynamical density functional theory \cite{EspanolL2009, AneroET2013, WittkowskiLB2012, WittkowskiLB2013, CamargoGdlTDZEDBC2018}, solid-state theory \cite{KakehashiF2004}, spin relaxation theory \cite{KivelsonO1974, Bouchard2007}, dielectric relaxation theory \cite{Khamzin2012,NigmatullinN2006}, spectroscopy \cite{SchmittSR2006}, calculation of correlation functions \cite{HansenMD2009}, plasma physics \cite{DiamondII2010}, and particle physics \cite{HuangKKR2011}. Due to its great importance, it is also studied in disciplines outside of physics such as mathematics \cite{Givon2005, ChorinHK2000, DominyV2017} and philosophy \cite{Wallace2015}. A significant extension of the formalism is therefore likely to have a strong impact on a large number of areas.

The currently used form of the Mori-Zwanzig formalism faces the problem that it cannot be directly applied to systems with time-dependent Hamiltonians. Those, however, are relevant for a lot of scenarios including soft matter systems subject to time-dependent external driving forces \cite{KomuraO2012, Menzel2015} or nuclear magnetic resonance (NMR) measurements with rapidly varying electromagnetic pulses \cite{Bouchard2007}. 

If arising, time-dependent Hamiltonians can sometimes be treated as additional external perturbations \cite{Grabert1982}. This requires, however, that the perturbation is sufficiently small and couples to the macroscopic variables only. Generalizations of the projection operator method towards non-Hamiltonian dynamical systems have been developed by Chorin, Hald, and Kupferman \cite{ChorinHK2000, ChorinHK2002}. Xing and Kim use mappings between dissipative and Hamiltonian systems \cite{Xing2010} to apply projection operators in the non-Hamiltonian case \cite{XingK2011}. The methods from Refs.\ \cite{ChorinHK2002, Xing2010, XingK2011}, however, are not applicable to quantum-mechanical systems. Moreover, approximation methods commonly applied in the context of the Mori-Zwanzig formalism, such as the linearization around thermodynamic equilibrium \cite{Grabert1982}, rely on the existence of a Hamiltonian.

The Mori-Zwanzig formalism exists in a variety of forms. The original theory developed by Mori \cite{Mori1965} can be applied to systems with time-independent Hamiltonians that are close to thermal equilibrium. A generalization towards systems far from equilibrium using time-dependent projection operators has been presented by Robertson \cite{Robertson1966}, Kawasaki and Gunton \cite{KawasakiG1973}, and Grabert \cite{Grabert1982, Grabert1978}. Recently, Bouchard has derived an extension of the Mori theory towards systems with time-dependent Hamiltonians \cite{Bouchard2007} that can be applied close to equilibrium. What is still missing, however, is a general formalism for systems that have a time-dependent Hamiltonian and are far from equilibrium.

General discussions of projection operators used for deriving coarse-grained equations of motion also exist for time-dependent Hamiltonians in nonequilibrium statistical physics \cite{UchiyamaS1999} and quantum field theory \cite{KoideM2000}. These methods, like the Bouchard equation \cite{Bouchard2007}, are derived without explicitly assuming a close-to-equilibrium situation and therefore always formally valid. However, they use time-independent projection operators and therefore do not form generalizations of the full Mori-Zwanzig formalism presented by Grabert \cite{Grabert1982}. The problem with time-independent projection operators is that they are only useful if the relevant dynamics is linear, since nonlinear effects are projected out together with the irrelevant dynamics \cite{Grabert1982,Zwanzig2001}. A useful description of far-from-equilibrium dynamics needs to be capable of representing nonlinear couplings. Moreover, these methods do not involve appropriate generalizations of the correlators, which are essential for the Mori-Zwanzig formalism.

In this article, we therefore derive an extension of the Mori-Zwanzig formalism for far-from-equilibrium systems with time-dependent Hamiltonians. For this purpose, we extend Grabert's treatment by introducing suitable generalized correlators and equations of motion. We also discuss how observables with explicit time dependence can be treated within this framework. Furthermore, we develop approximation methods that can be used to simplify the resulting equations. These approximations include a generalized Markovian approximation for slow variables, a linearization around thermodynamic equilibrium which recovers the Bouchard equation, the Magnus expansion for time-ordered exponentials, and the classical limit. To demonstrate how the formalism can be used, we apply it in combination with the approximation methods to the important case of spin relaxation. In this context, we show how the Bloch equations for spin relaxation in the presence of a time-dependent external magnetic field can be derived.

In the classical case, mappings between dissipative and Hamiltonian systems as described by Xing and Kim \cite{Xing2010, XingK2011} can extend the applicability of our formalism towards non-Hamiltonian systems, since they allow for the construction of a corresponding Hamiltonian. Since our method is not restricted to time-independent Hamiltonians, this would even allow for the treatment of arbitrary nonautonomous dynamical systems. It is also possible to include stochastic equations in this way. The stochastic contributions can be modeled using a harmonic bath Hamiltonian \cite{Xing2010, XingK2011, Zwanzig1973}.

This article is organized as follows: In \cref{derivation}, we derive the extended projection operator formalism. Approximations are discussed in \cref{approximations}. In \cref{spins}, the Bloch equations are obtained within our framework. Finally, we summarize our results in \cref{conclusions}.

\section{\label{derivation}Derivation of the extended projection operator formalism}
\subsection{\label{time}Time-dependent Liouvillians}
In the Heisenberg picture of quantum mechanics, a system is described by a set of observables $\{A_i(t)\}$ corresponding to time-dependent Hermitian Hilbert space operators. Any operator $A(t)$ obeys the Heisenberg equation of motion \cite{Muenster2010}
\begin{equation}
\tdif{}{t}A(t) = \frac{\ii}{\hbar}[\HH(t), A(t)] + \pdif{}{t} A(t),
\label{heisenberg}%
\end{equation}
where $\ii$ is the imaginary unit, $\hbar = h/(2\pi)$ the reduced Planck constant, and $[\cdot, \cdot]$ a commutator. With $\HH(t)$, we denote the Heisenberg picture Hamiltonian of the system, which can differ from the Schr{\"o}dinger picture Hamiltonian $\HS(t)$, if the Hamiltonian has explicit time dependence in the Schr\"odinger picture \cite{Muenster2010}. In most cases, one assumes that the observables are not explicitly time-dependent. Defining the Heisenberg picture Liouvillian\footnote{Some authors include the imaginary unit in the definition of the Liouvillian, so that \cref{liouville} reads $\frac{\dif}{\dif t}A(t) = \LH(t) A(t)$. In this case, the form of all resulting equations has to be modified accordingly. This has no additional effect on any of the calculations.}
\begin{equation}
\LH(t) = \frac{1}{\hbar}[\HH(t), \cdot],
\label{liouh}%
\end{equation}
we can write \cref{heisenberg} as
\begin{equation}
\frac{\dif}{\dif t}A(t) = \ii \LH(t) A(t).
\label{liouville}%
\end{equation}
The difference between Heisenberg picture Hamiltonian $\HH(t)$ and Schr\"odinger picture Hamiltonian $\HS(t)$ leads to a difference between the Heisenberg picture Liouvillian $\LH(t)$ and a Schr\"odinger picture Liouvillian $\LS(t)$ defined as
\begin{equation}
\LS(t) = \frac{1}{\hbar}[\HS(t), \cdot].
\label{lious}%
\end{equation}

If the Hamiltonian does not depend on time, the Liouvillian also does not and we denote them by $H$ and $L$, respectively. In the time-independent case, Schr\"odinger and Heisenberg picture Hamiltonians as well as Schr\"odinger and Heisenberg picture Liouvillians coincide and we do not need a subscript to distinguish between them. Equation \eqref{liouville} can then formally be solved as
\begin{equation}
A(t) = e^{\ii Lt} A_0,
\end{equation}
where $A_0 = A(0)$. However, this is no longer possible, if the Hamiltonian depends on time, since one usually\footnote{There are special situations in which, even though the Hamiltonian is time-dependent, the relevant Liouvillian is not, because the time-dependent part of the Hamiltonian commutes with the observable of interest, which is not true in general.} has a time-dependent Liouvillian then. In this case, \cref{liouville} has to be solved using time-ordered exponentials. 

For convenience, we repeat here the standard derivation presented, e.g., by Peskin and Schroeder \cite{PeskinS1995}. We assume $t > 0$. Integrating \cref{liouville} gives
\begin{equation}
A(t) = A_0 + \ii \INT{0}{t}{t'} \LH(t') A(t').
\end{equation}
This can be solved by iteration:
\begin{equation}
\begin{split}
A(t) &= A_0 + \ii \INT{0}{t}{t'} \LH(t') A_0 \\
&\quad\:\! + \ii^2 \INT{0}{t}{t'}\!\INT{0}{t'}{t''} \LH(t') \LH(t'') A_0 + \dotsb
\end{split}
\end{equation}
Since the Liouvillians stand in time order, where the operators on the left correspond to later times, we can use the identity \cite{PeskinS1995}
\begin{equation}
\begin{split}
&\INT{0}{t}{t_1}\!\INT{0}{t_1}{t_2} \dotsb \INT{0}{t_{n-1}}{t_n} \LH(t_1) \LH(t_2) \dotsb \LH(t_n)\\
&= \frac{1}{n!}\INT{0}{t}{t_1} \dotsb \INT{0}{t}{t_n} T_L(\LH(t_1) \dotsb \LH(t_n))
\end{split}
\label{timeorder}%
\end{equation}
with the left-time-ordering operator $T_L$. That operator is, for a time-dependent operator $\psi(t)$, defined as
\begin{equation}
T_L\psi(t_1)\psi(t_2) =
\begin{cases}
\psi(t_1)\psi(t_2), \quad \text{ if }t_1 > t_2,\\
\psi(t_2)\psi(t_1), \quad \text{ otherwise}
\end{cases}
\end{equation}
so that operators are always ordered in such a way that time increases from right to left.
For a $t > 0$, we arrive at the solution
\begin{equation}
\begin{split}
A(t) &= A_0 + \ii \INT{0}{t}{t'} \LH(t') A_0 \\
& \quad\, + \frac{\ii^2}{2} \INT{0}{t}{t'}\!\INT{0}{t}{t''} T_L 
\LH(t')\LH(t'') A_0 + \dotsb \\
&= T_L\bigg(\!\exp\!\bigg(\ii \INT{0}{t}{t'} \LH(t')\bigg)\!\bigg) A_0.
\end{split}\label{sol}%
\end{equation}
We also consider the backwards case, which we need further below. For a $t_0 > t$, we get
\begin{equation}
\begin{split}
A(t) &= A(t_0)+ \ii \INT{t_0}{t}{t'} \LH(t') A(t') \\
&= A(t_0) - \ii \INT{t}{t_0}{t'} \LH(t') A(t') \\
&= A (t_0) - \ii \INT{t}{t_0}{t'} \LH(t') A(t_0) \\
&\quad\, + (-\ii)^2 \INT{t}{t_0}{t'}\!\INT{t'}{t_0}{t''} \LH(t') \LH(t'') A(t_0) \\
&\quad\, + \dotsb
\end{split}
\end{equation}
As the Liouvillians are also time-ordered here, now with earlier times on the left, we obtain, using the identity
\begin{equation}
\begin{split}
&\INT{t}{t_0}{t_1}\!\INT{t_1}{t_0}{t_2} \dotsb \INT{t_{n-1}}{t_0}{t_n} \LH(t_1) \LH(t_2) \dotsb \LH(t_n)\\
&= \frac{1}{n!}\INT{t}{t_0}{t_1} \dotsb \INT{t}{t_0}{t_n} T_R(\LH(t_1) \dotsb \LH(t_n)),
\end{split}
\end{equation}
where $T_R$ is the right-time-ordering operator that puts later times on the right, the solution
\begin{equation}
A(t) = T_R \bigg(\!\exp\!\bigg(\!-\ii \INT{t}{t_0}{t'} \LH(t')\bigg)\!\bigg) A (t_0).
\label{back2}%
\end{equation}
In particular, this means (replace $t \to 0$ and $t_0 \to t > 0$)
\begin{equation}
A_0 = T_R \bigg(\!\exp\!\bigg(\!-\ii\INT{0}{t}{t'} \LH(t')\bigg)\!\bigg) A(t).
\label{back}%
\end{equation}
We introduce the abbreviations
\begin{align}
\exp_L(x) &:= T_L \exp(x),\\
\exp_R(x) &:= T_R \exp(x)
\end{align}
to simplify our notation in the following.

An important difference between left- and right-time-ordered exponentials is their behavior under differentiation. For left-time-ordered exponentials one has \cite{HolianE1985}
\begin{equation}
\frac{\dif}{\dif t} \exp_L\!\bigg(\ii \INT{0}{t}{t'} \LH(t')\bigg) \!
= \ii \LH(t) \exp_L\!\bigg(\ii \INT{0}{t}{t'} \LH(t')\bigg),
\end{equation}
i.e., the inner derivative stands on the left of the exponential. In contrast, for right-time-ordered exponentials one has
\begin{equation}
\frac{\dif}{\dif t} \exp_R\!\bigg(\ii \INT{0}{t}{t'} \LH(t')\bigg) \!
= \exp_R\!\bigg(\ii \INT{0}{t}{t'} \LH(t')\bigg)\ii \LH(t),
\end{equation}
where the inner derivative is on the right. This is relevant, because usually $\ii \LH(t)$ and the time-ordered exponential do not commute due to the noncommutativity of the Hamiltonians at different points in time. For some parts of the derivation of the extended projection operator formalism, it is essential that inner derivatives stand on the right. 
The identity
\begin{equation}
\exp_L\!\bigg(\ii \INT{0}{t}{t'} \LH(t')\bigg) = \exp_R\!\bigg(\ii \INT{0}{t}{t'} \LS(t')\bigg)
\label{hs}%
\end{equation}
allows to write \cref{sol} as 
\begin{equation}
A(t) = \exp_R\!\bigg(\ii \INT{0}{t}{t'} \LS(t')\bigg) A_0.
\label{anticausal}%
\end{equation}
Using the Schr{\"o}dinger picture Liouvillians therefore allows to take advantage of the properties of right-time-ordered exponentials. 

Equation \eqref{hs} can be proven in two ways. The first option is a direct calculation. Writing out the operator exponentials as expansions of commutators using the definitions \eqref{liouh} and \eqref{lious} for Heisenberg and Schr{\"o}dinger picture Liouvillians, respectively, and inserting the transformation rule
\begin{equation}
\HH(t) = U^\dagger(t)\HS(t)U(t)
\end{equation}
with the unitary operator \cite{PeskinS1995}
\begin{equation}
U(t) = \exp_L\!\bigg(\!- \frac{\ii}{\hbar} \INT{0}{t}{t'} \HS(t')\bigg)
\end{equation}
and a Hermitian adjoint denoted by $^\dagger$, gives, after sorting terms, the identity \eqref{hs}. Here, we use the second option, which is more elegant and gives better insights into the physics behind \cref{hs}. If we work in the Schr{\"o}dinger instead of the Heisenberg picture, the operators are time-independent, while the wave functions are time-dependent. This leads to a time-dependent density operator $\rho(t)$ \cite{JensenM1991}, since the density operator is constructed from the wave functions. The time evolution of the density operator is, in the Schr{\"o}dinger picture, given by the Liouville-von Neumann equation \cite{Grabert1982}
\begin{equation}
\frac{\dif}{\dif t} \rho(t) = - \ii \LS(t) \rho(t),
\label{vonneu}%
\end{equation}
which has the formal solution
\begin{equation}
\rho(t) = \exp_L\!\bigg(\! -\ii \INT{0}{t}{t'} \LS(t')\bigg)\rho(0).
\end{equation}
The mean value of an operator $A$ in the Schr{\"o}dinger picture is therefore obtained through \cite{Grabert1982}
\begin{equation}
\begin{split}
a(t) &= \Tr(\rho(t)A)\\
&= \Tr\!\bigg(\!\exp_L\!\bigg(\!-\ii \INT{0}{t}{t'} \LS(t')\bigg)\rho(0)A\bigg),
\end{split}\label{schroed}%
\end{equation}
where $\Tr$ denotes the trace. We now wish to transform \cref{schroed} to the Heisenberg picture. This is important, because in the Mori-Zwanzig formalism $\rho(t)$ is usually prescribed as an initial condition at $t=0$, but not known for larger $t$. Using \cref{timeorder,sol}, we can write \cref{schroed} as
\begin{equation}
\begin{split}
a(t) &= \Tr\!\bigg(\!\bigg(1 - \ii \INT{0}{t}{t'} \LS(t') \\
&\quad\,+(-\ii)^2 \INT{0}{t}{t'}\!\INT{0}{t'}{t''} \LS(t')\LS(t'') + \dotsb \bigg)\rho(0) A  \bigg)\\
&= \Tr(\rho(0)A) - \INT{0}{t}{t'} \Tr(\ii \LS(t')\rho(0)A) \\
&\quad\, +\INT{0}{t}{t'}\!\INT{0}{t'}{t''} \Tr(\ii \LS(t')\ii \LS(t'')\rho(0)A) + \dotsb
\end{split}\label{aentw}\raisetag{6em}%
\end{equation}
We can now use the relation \cite{Grabert1982}
\begin{equation}
\Tr(X \ii \LS Y) = - \Tr((\ii \LS X) Y),
\label{trace}%
\end{equation}
which is based on $\LS(t)^\dagger=-\LS(t)$. This relation is applied repeatedly to the right-hand side of \cref{aentw} (once to the second term, twice to the third term,\dots). The result is
\begin{equation}
\begin{split}
a(t) &= \Tr(\rho(0)A) + \INT{0}{t}{t'} \Tr(\rho(0)\ii \LS(t')A) \\
&\quad\,+ \INT{0}{t}{t'}\!\INT{0}{t'}{t''} \Tr(\rho(0)\ii \LS(t'')\ii \LS(t')A) + \dotsb\\
&= \Tr\!\bigg(\rho(0) \bigg(1 + \ii \INT{0}{t}{t'} \LS(t') \\
&\quad\,+\ii^2 \INT{0}{t}{t'}\!\INT{0}{t'}{t''} \LS(t'')\LS(t') + \dotsb\bigg)A \bigg)\\
&= \Tr\!\bigg(\rho(0) \exp_R\!\bigg( \ii \INT{0}{t}{t'} \LS(t')\bigg)A\!\bigg).
\end{split}\label{rto}%
\end{equation}
It shows that, if one transforms from the Schr{\"o}dinger to the Heisenberg picture, the Schr{\"o}dinger picture Liouvillians act on the operators \textit{in right time order}. On the other hand, we also could have worked directly in the Heisenberg picture. In this case, the mean value is given by \cite{Grabert1982}
\begin{equation}
\begin{split}
a(t) &= \Tr(\rho(0)A(t))\\
&= \Tr\!\bigg(\rho(0)\exp_L\!\bigg(\ii \INT{0}{t}{t'} \LH(t')\bigg)A\bigg),
\end{split}\label{heis}%
\end{equation}
where we have used \cref{sol}. Comparing \cref{rto,heis} gives
\begin{equation}
\exp_L\!\bigg(\ii \INT{0}{t}{t'} \LH(t')\bigg)A = \exp_R\!\bigg( \ii \INT{0}{t}{t'} \LS(t')\bigg)A.
\end{equation}
Since we have made no assumptions about the form of $A$, this also proves the identity (\ref{hs}). 
For extended discussions on transformations between Schr{\"o}dinger and Heisenberg pictures, see Holian and Evans for the classical \cite{HolianE1985, EvansM2008} and Uchiyama and Shibata \cite{UchiyamaS1999} for the quantum mechanical case.

\subsection{\label{poc}Projection operator and correlator}
As a starting point for our extension, we use the formalism presented by Grabert \cite{Grabert1978, Grabert1982}. Since it works in the nonlinear regime arbitrarily far from thermal equilibrium, it is by now the most general projection operator theory for systems with not explicitly time-dependent Hamiltonians. Extending this formalism will therefore allow to obtain one that has a greater range of applicability than existing theories.

We start with introducing a relevant probability density $\bar{\rho}(t)$. The complete microscopic state of the system is described by the actual probability density $\rho$, which is typically not known. Supposing that the macroscopic thermodynamic state of the system is well described by a set of macroscopic observables $\{A_i\}$ with mean values $\{a_i(t)\}$, one constructs the relevant density in such a way that it is only a function of the $\{a_i(t)\}$ and macroequivalent to the actual density $\rho(t)$ in the sense that \cite{Grabert1978}
\begin{equation}
\Tr(\rho(t) A_i) = \Tr(\bar{\rho}(t) A_i) = a_i(t).
\label{macroequivalence}%
\end{equation}
Although any choice meeting those restrictions is formally possible, the resulting equations are particularly useful, if the relevant density is a good approximation for the actual density. This is facilitated, if one can assume -- as we do for this derivation -- that the system is initially prepared in the state $\rho(0) = \bar{\rho}(0)$ \cite{Grabert1982}. Following Grabert \cite{Grabert1978} and Anero et al.\ \cite{AneroET2013}, we use the microcanonical form\footnote{Throughout this article, summation over each index appearing twice in a term is assumed.}
\begin{equation}
\bar{\rho}(t) = \frac{1}{Z(t)}e^{-\lambda_j(t) A_j},
\label{rhobar}%
\end{equation}
where $Z(t)$ is a normalization function ensuring $\Tr(\bar{\rho}(t))=1$. The conjugate variables $\{\lambda_i(t)\}$ are defined by the macroequivalence condition \eqref{macroequivalence} and therefore functions of the $\{a_i(t)\}$. In the case of non-Hermitian operators $A_i$, \cref{rhobar} has to be modified to ensure that the statistical operator is still Hermitian. Moreover, the macroequivalence condition \eqref{macroequivalence} can, in particular cases, not be fulfilled for non-Hermitian operators, if the form \eqref{rhobar} is chosen. However, this is unproblematic, since almost all applications are based on Hermitian operators. We will therefore, throughout this work, assume that all observables are Hermitian, if not stated otherwise. For a discussion of possible modifications of \cref{rhobar} in the case of certain important non-Hermitian operators, see Appendix \ref{appendixb}.

A motivation for choosing this form is that it maximizes the Gibbs entropy based on the available information \cite{Grabert1978,AneroET2013}, which here is given by the $\{a_i(t)\}$. If one defines a coarse-grained dimensionless entropy as
\begin{equation}
S(t) = - \Tr(\bar{\rho}(t) \ln(\bar{\rho}(t))),
\end{equation}
the conjugate variables are given by \cite{Grabert1978}
\begin{equation}
\lambda_i(t) = \frac{\partial S(t)}{\partial a_i(t)}. 
\label{lambda}%
\end{equation}
Typically, the use of the entropy as a thermodynamic potential is particularly appropriate for closed systems. It has, in the case of time-dependent Hamiltonians, the advantage that the Hamiltonian does not appear in its definition, so that it is not explicitly time-dependent. Alternatively, one can use a free energy combined with a canonical relevant density, which we discuss in \cref{bou}.

One might question whether a relevant probability density as in \cref{rhobar} is in fact a reasonable approximation for the actual probability density in the case of time-dependent Hamiltonians, since those systems are typically driven out of equilibrium, making arguments based on maximal-entropy principles seemingly less plausible. There are three reasons justifying to choose the form \eqref{rhobar}. First, the motivation for choosing the maximal Gibbs entropy form is that this is the \ZT{least biased} form regarding missing knowledge about the microscopic configuration, i.e., it is justified from an information theoretic point of view \cite{AneroET2013}. Second, the resulting equations of motion are exact regardless of the choice of $\bar{\rho}(t)$ as long as the macroequivalence condition \eqref{macroequivalence} is satisfied \cite{Grabert1978}, and this choice allows to express this condition in a useful way as equations \eqref{lambda} for the conjugate variables $\{\lambda_i(t)\}$. Third, we can assume that the time dependence (e.g., by an external field) is switched on at $t=0$, which means that the initial condition $\rho(0) = \bar{\rho}(0)$ is satisfied. For a detailed discussion of initial nonequilibrium states see Ref.\ \cite{Zwanzig2001}.

The aim is now to separate the dynamics into two parts: the \textit{organized motion}, which is entirely determined by the macroscopic mean values, i.e., the relevant density $\bar{\rho}(t)$, and the \textit{disorganized motion}, which corresponds to deviations of the actual dynamics from the organized motion. For analyzing this, we decompose the operator $A_i(t)$ as
\begin{equation}
A_i(t) = a_i(t) + \delta A_i(t),
\end{equation}
where $\delta A_i(t)$ describes the fluctuations of $A_i(t)$ around the mean value $a_i(t)$. The organized motion of the mean values is given by \cite{Grabert1978}
\begin{equation}
\bar{\dot{a}}_i(t) = \Tr(\bar{\rho}(t) \ii \LS(t)A_i) =: v_i(t).
\label{v}%
\end{equation} 
This relation defines the \textit{organized drift} $v_i(t)$ \cite{Grabert1978}. Since the Liouvillian acts directly on the Schr{\"o}dinger picture operator $A_i$, we have to use the Schr{\"o}dinger picture Liouvillian $\LS(t)$. Note that, for a time-dependent Liouvillian as assumed here, one cannot unambiguously write $\dot{A} = \ii \LS A$ as it is done in the usual presentations. The deviation
\begin{equation}
\ww_i(t) = \dot{a}_i(t) - v_i(t)
\label{DisorganizedDrift}%
\end{equation}
is the \textit{disorganized drift}. Based on the identity \cite{Grabert1978}
\begin{equation}
- \ii \LS(t) \bar{\rho}(t) = \lambda_j(t) \INT{0}{1}{\alpha} e^{-\alpha \lambda_k(t) A_k} \ii \LS(t) A_j e^{\alpha \lambda_l(t) A_l}\bar{\rho}(t)
\label{lidentity}%
\end{equation}
and the fact that the time evolution of the relevant density is given by
\begin{equation}
\pdif{}{t}\bar{\rho}(t) = \frac{\partial \bar{\rho}(t)}{\partial a_j(t)}\dot{a}_j(t),
\end{equation}
Grabert shows that the organized motion of the fluctuations $\delta A_i(t)$, which is defined by demanding that it does not lead to deviations of $\rho(t)$ from $\bar{\rho}(t)$, is given by \cite{Grabert1978}
\begin{equation}
\delta \bar{\dot{A}}_i(t) = \frac{\partial v_i(t)}{\partial a_j(t)}\delta A_j(t) = \Omega_{ij}(t) \delta A_j(t)
\label{linear}%
\end{equation}
with the frequency matrix
\begin{equation}
\Omega_{ij}(t) = \pdif{v_i(t)}{a_j(t)}.
\label{Omegaij}%
\end{equation}

Up to now, our treatment of the organized motion did not differ from Grabert's treatment, since none of the steps done so far hangs on the fact that the time evolution is given by $X(t) = e^{\ii Lt}X$. New and interesting aspects come into play, however, if we introduce the projection operator $P(t)$. The aim of the projection operator is to extract the organized motion from the total dynamics, which means that the effect of the projection operator is determined by the previously derived relations for the organized motion. Grabert uses these results to write \cite{Grabert1978}
\begin{equation}
\begin{split}
\bar{\dot{A}}_i(t) &= \bar{\dot{a}}_i(t) + \delta \bar{\dot{A}}_i(t)\\
&= \Tr(\bar{\rho}(t) \dot{A}_i) + \delta A_j(t)\Tr\!\bigg(\frac{\partial \bar{\rho}(t)}{\partial a_j(t)}\ii L A_i\bigg)\\
&= e^{\ii Lt} P(t)\ii L A_i,
\end{split}
\end{equation}
where the \textit{projection operator} is given by
\begin{equation}
P(t)X = \Tr(\bar{\rho}(t)X) + (A_j - a_j(t))\Tr\!\bigg(\frac{\partial \bar{\rho}(t)}{\partial a_j(t)}X\bigg).
\label{projection}%
\end{equation}
If the relevant variables are not explicitly time-dependent, $P(t)X$ has the projection operator property\footnote{For explicitly time-dependent operators, this still holds for $t = t'$, so that $P(t)$ in \cref{projection} is a projection operator.} \cite{Grabert1982,Grabert1978}
\begin{equation}
P(t)P(t')X = P(t')X.
\end{equation}
It is important here that $P(t)X$ is a Schr{\"o}dinger picture operator that is then propagated using $e^{\ii Lt}$ to get the value of the projected observable at time $t$. Therefore, in order to be able to directly apply the projection operator \eqref{projection} used by Grabert, we need to use Schr{\"o}dinger picture Liouvillians and, therefore, right-time-ordered exponentials.

An important part of the theory -- and another point where there is a crucial difference between our formalism and the standard procedure -- is the choice of a suitable correlator that is needed for specifying correlation functions. This correlator is of particular interest here as it provides a generalization of the scalar product in the space of dynamical variables. Grabert introduces the \textit{generalized canonical correlation} \cite{Grabert1978}
\begin{equation}
\begin{split}
(X(t), Y(s)) &= \INT{0}{1}{\alpha} \Tr\!\big(\bar{\rho}(s)(e^{-\ii Ls}X(t))e^{-\alpha \lambda_j(s) A_j} \\
&\qquad\qquad\quad\,\:\! (e^{-\ii Ls}Y^\dagger(s)) e^{\alpha \lambda_k(s)A_k}\big)
\end{split}\label{general}\raisetag{5ex}%
\end{equation}
for two time-dependent operators $X(t)$ and $Y(s)$. This correlator is used to derive certain relations, in particular an equation for the frequency matrix \cite{Grabert1978}
\begin{equation}
\Omega_{ij}(t) = (\delta A_j(t), \delta A_k(t))^{-1}(\dot{A}_i(t), \delta A_k(t)).
\label{relation}%
\end{equation}
The generalized canonical correlation \eqref{general} reduces to Kubo's correlation functions in thermal equilibrium and to the standard correlation functions of statistical mechanics for classical systems. Its main purpose in the Mori-Zwanzig formalism is that the projection operator can be written in terms of it. In the general case, however, \cref{general} does not have the properties of a usual correlator. It has this specific form precisely because this allows to derive \cref{relation}. An appropriate generalization of the canonical correlator towards time-dependent Liouvillians therefore has to be constructed in such a way that \cref{relation} still holds. We found that the best choice for the \textit{generalized correlator}, i.e., the simplest form that meets the requirement \eqref{relation} and reduces to \cref{general} for time-independent Liouvillians, is
\begin{equation}
\begin{split}
&(X(t), Y(s))\\ 
&\!=\INT{0}{1}{\alpha} \Tr\!\bigg(\bar{\rho}(s)\\ 
&\qquad\quad\;\;\:\! \bigg(\exp_L\!\bigg(\!-\ii \INT{0}{s}{t'} \LS(t')\bigg) X(t)\bigg)e^{-\alpha \lambda_j(s) A_j} \\
&\qquad\quad\;\;\:\! \bigg(\exp_L\!\bigg(\!-\ii \INT{0}{s}{t'} \LS(t')\bigg) Y^\dagger(s)\bigg) e^{\alpha \lambda_k(s)A_k}\bigg).
\end{split}
\label{correlation}\raisetag{14ex}%
\end{equation}
This correlator, together with the fact that is should be written using Schr{\"o}dinger picture Liouvillians, constitutes the first main result of this section. In Appendix \ref{appendixa}, we explicitly prove that this correlator gives the desired result for the case of Hermitian operators, which correspond to physical observables. Since \cref{correlation} reduces to \cref{general} for a time-independent Liouvillian, it has the other properties of a correlator mentioned above.

The correlator \eqref{correlation} can be used to re-write the equation \eqref{projection} for the projection operator $P(t)$. With \cref{iden2,partialal} (see Appendix \ref{appendixa}), one can write 
\begin{equation} 
\begin{split}
&\:\! \Tr\!\bigg(\frac{\partial \bar{\rho}(t)}{\partial a_j(t)} X\bigg)\\
&= \Tr\!\bigg(\frac{\partial \bar{\rho}(t)}{\partial \lambda_k(t)}\frac{\partial \lambda_k(t)}{\partial a_j(t)} X\bigg)\\
&= - (\delta A_j(t), \delta A_k(t))^{-1} \Tr\!\bigg(\frac{\partial \bar{\rho}(t)}{\partial \lambda_k(t)} \\
&\qquad \exp_L\!\bigg(\!-\ii \INT{0}{t}{t'} \LS(t')\bigg) \exp_R\!\bigg(\ii \INT{0}{t}{t'} \LS(t')\bigg)X\bigg)\\
&= (\delta A_j(t), \delta A_k(t))^{-1}\bigg(\exp_R\!\bigg(\ii \INT{0}{t}{t'} \LS(t')\bigg)X, \delta A_k(t)\bigg).
\end{split}\raisetag{18ex}%
\end{equation}
Therefore, \cref{projection} can be re-written as
\begin{equation}
\begin{split}
P(t)X &= \Tr(\bar{\rho}(t)X) \\
&\quad\,+ (A_j - a_j(t))(\delta A_j(t), \delta A_k(t))^{-1}\\
&\quad\quad\; \bigg(\exp_R\!\bigg(\ii \INT{0}{t}{t'} \LS(t')\bigg)X, \delta A_k(t)\bigg).
\end{split}\label{projectioncorr}%
\end{equation}

\subsection{\label{DerivationEOMs}Time evolution of relevant variables}
To  obtain an equation of motion for the relevant dynamics, we follow the familiar procedure of decomposing the time-evolution operator used, e.g., by Zwanzig \cite{Zwanzig2001}, Grabert \cite{Grabert1982}, and Bouchard \cite{Bouchard2007}. We use right-time-ordered exponentials due to the properties they have if they are acted upon by the time derivative (see \cref{time}). To be able to do this, we also use Schr{\"o}dinger picture Liouvillians.

We start with
\begin{equation}
\begin{split}
\exp_R\!\bigg(\ii \INT{0}{t}{t'} \LS(t')\bigg) \! &= \exp_R\!\bigg(\ii \INT{0}{t}{t'} \LS(t')\bigg) \\
&\quad\; (P(t) + Q(t)),
\end{split}
\end{equation}
where $Q(t) = \id - P(t)$ is the operator complementary to $P(t)$. From the term
\begin{equation}
\WW(t) = \exp_R\!\bigg(\ii \INT{0}{t}{t'} \LS(t')\bigg)Q(t)
\end{equation}
we get the time derivative
\begin{equation}
\begin{split}
\dot{\WW}(t) &= \exp_R\!\bigg(\ii \INT{0}{t}{t'} \LS(t')\bigg)
\big(\ii \LS(t)Q(t) - \dot{P}(t)\big) \\
&= \WW(t)\ii \LS(t)Q(t)\\ 
&\quad\:\! + \exp_R\!\bigg(\ii \INT{0}{t}{t'} \LS(t')\bigg) \big( P(t) \ii \LS(t)Q(t) - \dot{P}(t)\big).
\end{split}\label{bdot}\raisetag{10ex}%
\end{equation} 
The last term of \cref{bdot} depends on macroscopic variables and can be considered a known inhomogeneity. 
Realizing that the homogeneous solution of \cref{bdot} is $\WW(t) = \WW(u) G(u,t)$ with the time-ordered exponential
\begin{equation}
G(s,t) = \exp_R\!\bigg(\ii \INT{s}{t}{t'} \LS(t')Q(t')\bigg)
\end{equation}
and using the initial condition
\begin{equation}
\WW(u) = \exp_R\!\bigg(\ii \INT{0}{u}{t'} \LS(t')\bigg)Q(u),
\end{equation}
we thus find the result \cite{Grabert1982}
\begin{equation}
\begin{split}
\WW(t) &= \exp_R\!\bigg(\ii \INT{0}{u}{t'} \LS(t')\bigg) Q(u) G(u,t) \\
&\quad +\INT{u}{t}{s} \exp_R\!\bigg(\ii \INT{0}{s}{t'} \LS(t')\bigg) \\
&\qquad\qquad\;\, \big(P(s)\ii \LS(s)Q(s) - \dot{P}(s)\big) G(s,t).
\end{split}
\end{equation}
This gives the operator identity
\begin{equation}
\begin{split}
&\:\! \exp_R\!\bigg(\ii \INT{0}{t}{t'} \LS(t')\bigg) \\ 
&= \exp_R\!\bigg(\ii \INT{0}{t}{t'} \LS(t')\bigg) P(t) \\
& \quad + \exp_R\!\bigg(\ii \INT{0}{u}{t'} \LS(t')\bigg) Q(u) G(u,t) \\
& \quad + \INT{u}{t}{s} \exp_R\!\bigg(\ii \INT{0}{s}{t'} \LS(t')\bigg) \\
& \qquad\qquad\;\, \big(P(s) \ii \LS(s)Q(s) - \dot{P}(s)\big) G(s,t).
\end{split}\label{generaldyson}\raisetag{11ex}%
\end{equation}
If we consider the special case $u=0$ and a time-independent projection operator $P$, we can write this as\footnote{$G$ is a propagator acting on the subspace orthogonal to the relevant variables, so it commutes with $Q$.}
\begin{equation}
\begin{split}
&\:\! \exp_R\!\bigg(\ii \INT{0}{t}{t'} \LS(t')\bigg) Q\\ &=  \exp_R\!\bigg(\ii \INT{0}{t}{t'} \LS(t')Q\bigg) Q \\ 
& \quad \:\! + \INT{0}{t}{s} \exp_R\!\bigg(\ii \INT{0}{s}{t'} \LS(t')\bigg) P\\
& \qquad\qquad\;\;\:\! \ii \LS(s) \exp_R\!\bigg(\ii \INT{s}{t}{t'} \LS(t')Q\bigg)Q.
\end{split}\raisetag{11ex}%
\end{equation}
Comparing the prefactors of $Q$ on both sides of the equation gives\footnote{Of course, $AQ = BQ$ does in general not prove $A = B$, if $Q$ is a projection operator, since $A$ and $B$ could in principle have different effects on the subspace orthogonal to $Q$. However, since the choice of relevant observables determining the projection operator is free, we consider the case $Q = \id$, completing the proof. Alternatively, \cref{dyson} can be easily confirmed by taking the time derivative on both sides of the equation \cite{Bouchard2007}.}
\begin{equation}
\begin{split}
&\:\! \exp_R\!\bigg(\ii \INT{0}{t}{t'} \LS(t')\bigg) \\
&= \exp_R\!\bigg(\ii \INT{0}{t}{t'} \LS(t')Q\bigg) \\ 
&\quad\:\! + \INT{0}{t}{s} \exp_R\!\bigg(\ii \INT{0}{s}{t'} \LS(t')\bigg) P\\
& \qquad\qquad\;\;\:\! \ii \LS(s)\exp_R\!\bigg(\ii \INT{s}{t}{t'} \LS(t')Q\bigg).
\end{split}\label{dyson}\raisetag{11ex}%
\end{equation}
This is the generalization of the \textit{Dyson decomposition} derived by Holian and Evans \cite{HolianE1985, EvansM2008}. It reduces to the usual Dyson decomposition, if we assume that also the Liouvillian does not depend on time. In the case of time-independent projection operators, the generalized Dyson decomposition \eqref{dyson} can be used to derive the equation of motion more directly \cite{Bouchard2007}.

Applying the identity \eqref{generaldyson} to $\ii \LS(t) A_i$ gives
\begin{equation}
\begin{split}
\dot{A}_i(t) &= \exp_R\!\bigg(\ii \INT{0}{t}{t'} \LS(t')\bigg) P(t) \ii \LS(t) A_i\\
& \quad + \exp_R\!\bigg(\ii \INT{0}{u}{t'} \LS(t')\bigg) Q(u) G(u,t)\ii \LS(t) A_i\\
& \quad + \INT{u}{t}{s} \exp_R\!\bigg(\ii \INT{0}{s}{t'} \LS(t')\bigg) \\
& \qquad\qquad\;\, \big(P(s)\ii \LS(s)Q(s) - \dot{P}(s)\big) G(s,t)\ii \LS(t) A_i.
\end{split}\label{eqm}\raisetag{11ex}%
\end{equation}
Using \cref{projection} and \cite{Grabert1978}
\begin{equation}
\dot{P}(t) X = (A_j - a_j(t)) \dot{a}_k(t) \Tr\!\bigg(\frac{\partial^2 \bar{\rho}(t)}{\partial a_j(t) \partial a_k(t)} X\bigg)
\end{equation}
as well as introducing the after-effect function
\begin{equation}
K_i(t,s) = \Tr\!\big(\bar{\rho}(s) \ii \LS(s)Q(s)G(s,t) \ii \LS(t) A_i\big),
\label{aftereffect}%
\end{equation}
the memory function
\begin{equation}%
\begin{split}%
\phi_{ij}(t,s) &= \Tr\!\bigg(\frac{\partial \bar{\rho}(s)}{\partial a_j(s)} \ii \LS(s)Q(s)G(s,t) \ii \LS(t) A_i\bigg) \\
&\quad\, - \dot{a}_k(s)\Tr\!\bigg(\frac{\partial^2 \bar{\rho}(s)}{\partial a_j(s) \partial a_k(s)} G(s,t)\ii \LS(t) A_i\bigg),
\end{split}\label{phiij}\raisetag{9ex}%
\end{equation}%
and the random force
\begin{equation}
F_i(t,s) = \exp_R\!\bigg(\ii \INT{0}{s}{t'} \LS(t')\bigg) Q(s) G(s,t)\ii \LS(t) A_i,
\end{equation}
we can write \cref{eqm} as the equation of motion
\begin{equation}
\begin{split}
\dot{A}_i(t) &= v_i(t) + \Omega_{ij}(t)\delta A_j(t) \\
&\quad\:\! + \INT{u}{t}{s} \big(K_i(t,s) + \phi_{ij} (t,s)\delta A_j(s)\big) \\
&\quad\:\! + F_i(t,u).
\end{split}\label{eqofmotion}%
\end{equation}
As for the usual Mori-Zwanzig formalism, averaging of \cref{eqofmotion} allows to derive the dynamics of the mean values \cite{Grabert1982}
\begin{equation}
\dot{a}_i(t) = v_i(t) + \INT{u}{t}{s} K_i(t,s) + f_i(t,u)
\label{average}%
\end{equation}
with $f_{i}(t,u) = \Tr (\rho(0) F_i(t,u))$. The term $f_i(t,u)$ can be dropped, if one assumes that the initial distribution $\rho(0)$ is given by the relevant density $\bar{\rho}(0)$ \cite{Grabert1982, Grabert1978}. Equations \eqref{eqofmotion} and \eqref{average} also lead to \cite{Grabert1982}
\begin{equation}
\begin{split}
\delta \dot{A}_i(t) &= \Omega_{ij}(t) \delta A_j(t) + \INT{u}{t}{s} \phi_{ij} (t,s) \delta A_j(s) \\
&\quad\:\!+ \delta F_i(t,u)
\end{split}\label{deltaa}%
\end{equation}
with $\delta F_i(t,u) = F_i(t,u) - f_i(t,u)$.

Equation \eqref{eqofmotion} is valid for any $u \in [0,t]$. The physical significance of $u$ is that the memory functions $K_i(t,s)$ and $\phi_{ij} (t,s)$ are functionals of the mean paths $\{a_i(s) : s \in [u,t]\}$. We get closed equations of motion \eqref{average} for the mean values, where $v_i(t)$ depends only on the current macroscopic state given by $a_i(t)$, while the memory functions depend on the macroscopic state at previous times in the interval $[u,t]$ \cite{Grabert1982}. Note that the validity of \cref{eqofmotion} does not rely on choosing the form \eqref{rhobar} for the relevant density $\bar{\rho}(t)$ as long as it has no explicit time dependence, but only on the form \eqref{projection} of the projection operator.

\subsection{\label{TimeEvolutionCij}Time evolution of correlation functions}
Using the previous results, it is possible to obtain very directly the rules for the time evolution of correlation functions defined via the generalized correlator \eqref{correlation}. This is done by taking the time derivative of the equation for the correlation function one is interested in and applying the rules obtained so far.

It is particularly interesting to consider the time correlation of the fluctuations, since here a closed and elegant result can be found. This correlation function is defined as
\begin{equation}
C_{ij}(t,u) = (\delta A_i(t), \delta A_j(u)).
\label{Cij}%
\end{equation}
Inserting \cref{deltaa} into the time derivative
\begin{equation}
\tdif{}{t}C_{ij}(t,u) = \tdif{}{t} (\delta A_i(t), \delta A_j(u)) 
= (\delta \dot{A}_i(t), \delta A_j(u))
\end{equation}
results in
\begin{equation}
\tdif{}{t}C_{ij}(t,u) = \Omega_{ik}(t) C_{kj}(t,u) + \INT{u}{t}{s} \phi_{ik}(t,s) C_{kj}(s,u),
\end{equation}
where we have used \cite{Grabert1978}
\begin{equation}
(\delta F_i(t,s), \delta A_j(s)) = 0.
\end{equation}

\subsection{\label{explicit}Explicitly time-dependent operators}
An interesting scenario that receives very little attention in the literature is that of explicitly time-dependent operators, for which the partial derivative with respect to $t$ in \cref{heisenberg} cannot be dropped. This scenario can occur, e.g., when the energy density is a relevant variable, since the corresponding operator adopts the explicit time dependence of the Hamiltonian. Therefore, we need to adapt our formalism to the full Heisenberg equation, if we want to derive equations of motion for explicitly time-dependent operators. We use the definition
\begin{equation}
P(t)X = \Tr(\bar{\rho}(t)X) + (\ASf{j}(t) - a_j(t))\Tr\!\bigg(\frac{\partial \bar{\rho}(t)}{\partial a_j(t)}X \bigg)
\end{equation}
for the projection operator to ensure that
\begin{equation}
P(t)\ASf{i}(t) = \ASf{i}(t)
\end{equation}
holds, where $\ASf{i}(t)$ is the explicitly time-dependent Schr{\"o}dinger picture operator corresponding to $A_i$ in \cref{projection}. The crucial question here is how to calculate the time evolution using Schr{\"o}dinger picture Liouvillians. In the Schr{\"o}dinger picture, the time evolution is given by
\begin{equation}
\dot{a}(t) = \Tr\!\bigg(\! \exp_L\!\bigg(-\ii \INT{0}{t}{t'} \LS(t')\bigg)\rho(0)\bigg(\ii \LS(t) + \pdif{}{t}\bigg)\AS(t)\bigg).
\label{schrodingertd}%
\end{equation}
Equation \eqref{schrodingertd} holds, because the explicit time dependence of $\AS(t)$ has no influence on the time evolution of the density operator $\rho(t)$. Using the procedure described in \cref{time}, a transformation to the Heisenberg picture gives
\begin{equation}
\dot{a}(t) = \Tr\!\bigg(\rho(0)\exp_R\!\bigg(\ii \INT{0}{t}{t'} \LS(t')\bigg)\!\bigg(\ii \LS(t) + \pdif{}{t}\bigg)\AS(t)\bigg).
\end{equation}
Therefore, the appropriate equation is
\begin{equation}
A(t) = \exp_R\!\bigg(\ii \INT{0}{t}{t'} \LS(t')\bigg)\AS(t),
\end{equation}
i.e., the partial derivative $\partial/\partial t$ does not have to be taken into account when constructing the exponential. The time derivative is then
\begin{equation}
\dot{A}(t) = \exp_R\!\bigg(\ii \INT{0}{t}{t'} \LS(t')\bigg)\!\bigg(\ii \LS(t) + \pdif{}{t}\bigg)\AS(t).
\label{exptd}%
\end{equation}
For deriving the equations of motion, we can apply the identity \eqref{generaldyson} for the time-ordered exponential in \cref{exptd}. A further difficulty here is that explicitly time-dependent observables also lead to explicitly time-dependent relevant densities and projection operators, so that
\begin{equation}
\begin{split}%
\dot{P}(t) X &= (\ASf{j}(t) - a_j(t)) \dot{a}_k(t) \Tr\!\bigg(\frac{\partial^2 \bar{\rho}(t)}{\partial a_j(t) \partial a_k(t) }X\bigg)\\
&\quad\:\! + \Tr\!\bigg(\pdif{\bar{\rho}(t)}{t} X\bigg) \\
&\quad\:\! + \bigg(\pdif{}{t}\ASf{j}(t)\bigg) \Tr\!\bigg(\frac{\partial \bar{\rho}(t)}{\partial a_j(t)}X\bigg)\\
&\quad\:\! + (\ASf{j}(t) - a_j(t))\Tr\!\bigg(\frac{\partial^2 \bar{\rho}(t)}{\partial t\partial a_j(t)}X\bigg).
\end{split}\raisetag{14ex}%
\end{equation}
The result is the equation of motion 
\begin{equation}
\begin{split}
\dot{A}_i(t) &= v_i(t) + \Omega_{ij}(t)\delta A_j(t) \\
&\quad\, + \INT{u}{t}{s} \big(K_i(t,s) + \phi_{ij} (t,s)\delta A_j(s) \\
&\qquad\qquad\;\;\, - \Ef{ij}(t,s)\partial^{\mathrm{ex}}_s \ASf{j}(s)\big) \\
&\quad\, + F_i(t,u)
\end{split}\label{eqofmotio}%
\end{equation}
with the (modified) definitions
\begin{gather}
\begin{split}
v_i(t) = \Tr\!\bigg(\bar{\rho}(t) \bigg(\ii \LS(t) + \pdif{}{t}\bigg) \ASf{i}(t)\bigg),
\end{split}\\
\begin{split}
K_i(t,s) &= \Tr\!\bigg(\bar{\rho}(s) \ii \LS(s)Q(s)G(s,t) \bigg(\ii \LS(t) + \pdif{}{t}\bigg) \ASf{i}(t)\\
&\qquad\;\;\; - \pdif{\bar{\rho}(s)}{s} G(s,t)\bigg(\ii \LS(t) + \pdif{}{t}\bigg) \ASf{i}(t)\bigg),
\end{split}\raisetag{7ex}\\
\begin{split}%
&\phi_{ij}(t,s)\\
&\!\!\:\!=\Tr\!\bigg(\frac{\partial \bar{\rho}(s)}{\partial a_j(s)} \ii \LS(s)Q(s)G(s,t) \bigg(\ii \LS(t) + \pdif{}{t}\bigg) \ASf{i}(t)\bigg) \\
&\!\quad -\Tr\!\bigg(\!\bigg(\dot{a}_k(s)\frac{\partial^2 \bar{\rho}(s)}{\partial a_k(s) \partial a_j(s)} + \frac{\partial^2 \bar{\rho}(s)}{\partial s \partial a_j(s)}\bigg) \\ 
&\!\qquad\qquad\!\:\! G(s,t)\bigg(\ii \LS(t) + \pdif{}{t}\bigg) \ASf{i}(t)\bigg),
\end{split}\raisetag{10ex}\\%
\begin{split}%
\Ef{ij}(t,s) = \Tr\!\bigg(\frac{\partial \bar{\rho}(s)}{\partial a_j(s)}G(s,t)\bigg(\ii \LS(t) + \pdif{}{t}\bigg) \ASf{i}(t)\bigg),
\end{split}\\
\begin{split}%
\partial^{\mathrm{ex}}_s \ASf{j}(s) = \exp_R\!\bigg(\ii \INT{0}{s}{t'} \LS(t')\bigg) \pdif{}{s} \ASf{j}(s),
\end{split}\\
\begin{split}
&F_i(t,s) \\
&\! = \exp_R\!\bigg(\ii \INT{0}{s}{t'} \LS(t')\bigg) Q(s) G(s,t)\bigg(\ii \LS(t) + \pdif{}{t}\bigg) \ASf{i}(t).
\end{split}\raisetag{7ex}%
\end{gather}
Interestingly, the explicitly time-dependent operators require an additional memory function $\Ef{ij}(t,s)$. The general form \eqref{eqofmotio} for the equation of motion is the second main result of this section. Note that here
\begin{equation}
\Omega_{ij}(t) = \frac{\partial v_i(t)}{\partial a_j(t)} \neq (\delta A_j(t), \delta A_k(t))^{-1}(\dot{A}_i(t), \delta A_k(t)),
\end{equation}
as the derivation in Appendix \ref{appendixa} assumes $\dot{A}(t) = \exp_R(\ii \TINT{0}{t}{t'} \LS(t')) \ii \LS(t)A$.

\subsection{Discussion}
The general form of our equations is very similar to Grabert's results for time-independent Hamiltonians \cite{Grabert1978, Grabert1982}. This is not surprising, since his treatment forms the starting point for our derivation. However, our discussion reveals some nontrivial new aspects. In particular, it is helpful to calculate the time evolution using Schr\"odinger picture Liouvillians, since this allows for the utilization of right-time-ordered exponentials. As discussed by Uchiyama and Shibata \cite{UchiyamaS1999}, right-time-ordered exponentials are a typical feature of Heisenberg picture projection operator formalisms. Moreover, we have constructed an equation for the generalization of the correlator, which also can be constructed based on Schr\"odinger picture Liouvillians, and explicitly calculated that it has the role it is supposed to have. What is completely new in treatments of the Mori-Zwanzig formalism is that we discuss the possibility of observables with explicit time dependence, to which we extend the formalism in a very natural way.

\section{\label{approximations}Approximations for the extended formalism}
The transport equations derived in \cref{derivation} are, in practical applications, typically too complex to be solved exactly. In this section, we therefore derive approximation methods that can be used to simplify the results in the important cases of slow variables, close-to-equilibrium systems and classical dynamics. Throughout this section, we assume that the observables do not have explicit time dependence. A generalization using the methods described in \cref{explicit} is straightforward. Without loss of generality, we assume that $u=0$, where $u$ is the reference time for the equation of motion \eqref{eqofmotion} introduced in \cref{DerivationEOMs}.

\subsection{\label{GeneralizedMarkovianApproximation}Generalized Markovian approximation}
In many situations, one is interested in slowly varying macroscopic variables. A typical example is the case of physical quantities that follow a conservation law \cite{Forster1989}. Slow variables allow the use of certain approximations, which simplify the structure of the resulting equations and facilitate practical applications. 
This is known as \textit{Markovian approximation}. 
In the following, we give a generalization of the corresponding discussion by Grabert \cite{Grabert1982} towards the case of time-dependent Liouvillians. 

Equations \eqref{trace} and \eqref{lidentity} allow to write the after-effect function \eqref{aftereffect} as
\begin{equation}
K_i(t,s) = R_{ij}(t,s)\lambda_j(s),
\label{AfterEffectFunction}%
\end{equation}
where we introduce the retardation matrix
\begin{equation}
\begin{split}
R_{ij}(t,s) &= \INT{0}{1}{\alpha} \Tr\!\big(\bar{\rho}(s) e^{\alpha \lambda_l(s)A_l} Q(s) \\
&\qquad\qquad\quad\;G(s,t)\ii \LS(t) A_i e^{-\alpha \lambda_k(s)A_k}\ii \LS(s) A_j\big).
\end{split}\label{RetardationMatrix}\raisetag{3em}%
\end{equation}
Modifications of the form of $R_{ij}(t,s)$ can be necessary for the case of non-Hermitian operators, since Eq. \eqref{RetardationMatrix} is based on Eq. \eqref{rhobar}. With \cref{AfterEffectFunction}, one can write \cref{average} as
\begin{equation}
\dot{a}_i(t) = v_i(t) + \INT{0}{t}{s} R_{ij}(t,s)\lambda_j(s).
\label{eom}%
\end{equation}
If we are interested in slow variables, we can drop terms that are of higher than second order in $\ii \LS(t) A$. Note that the assumption that $\ii \LS(t) A$ is small does not imply that $\ii \LS(t)$ is slowly varying. Since the elements $R_{ij}$ are of second order and variations of $a$ are of first order in $\ii \LS(t) A$, we can use the quasi-stationary path $a(s) = a(t)$ for $s \in [0,t]$ in the last term of \cref{eom} \cite{Grabert1982}. The reason is that a Taylor expansion gives
\begin{equation}
\begin{split}
a(s) &= a(t) + \dot{a}(t)(s-t) + \dotsb\\
&= \Tr(\rho(t) A) + \Tr(\rho(t)\ii \LS(t) A)(s-t) + \dotsb,
\end{split}\label{taylor}%
\end{equation}
i.e., any deviation of $a(s)$ from $a(t)$ is at least of first order in $\ii \LS(t)A$. For the same reason, we can replace $\lambda_j(s)$ by $\lambda_j(t)$, $\bar{\rho}(s)$ by $\bar{\rho}(t)$, and $P(s)$ by $P(t)$, noting that the relevant density and the projection operator are functionals of the mean values. No such approximation is possible, however, where the time dependence does not arise through the mean values of the (slow) relevant variables but directly through the Hamiltonian, i.e., we cannot replace $\ii \LS(s)$ by $\ii \LS(t)$. The time-ordered exponential $G(s,t)$ can, in the case of a time-independent Hamiltonian, be approximated as
\begin{equation}
\begin{split}
G(s,t) &= \exp_R\!\bigg(\ii \INT{s}{t}{t'} \LS Q(t')\bigg) 
\approx \exp_R\!\bigg(\ii \INT{s}{t}{t'} \LS Q(t)\bigg) \\
&= e^{\ii \LS(t-s)Q(t)} 
\approx e^{\ii \LS(t-s)},
\end{split}\raisetag{4ex}%
\end{equation}
where one uses the quasi-stationary approximation and the fact that $\ii \LS(t)P(t)$ is of order $\ii \LS(t) A$, since
\begin{equation}
\begin{split}
\ii \LS(t) P(t)X &= \Tr(\bar{\rho}(t)X)\ii \LS(t) \id \\ 
&\quad\:\! + (\ii \LS(t) A_j - a_j(t) \ii \LS(t) \id)\Tr\!\bigg(\frac{\partial \bar{\rho}(t)}{\partial a_j(t)}X\bigg)\\
& = \ii \LS(t) A_j\Tr\!\bigg(\frac{\partial \bar{\rho}(t)}{\partial a_j(t)}X\bigg).
\end{split}\raisetag{6ex}%
\end{equation}
In our case, however, we have
\begin{equation}
\begin{split}
G(s,t) &= \exp_R\!\bigg(\ii \INT{s}{t}{t'} \LS(t')Q(t')\bigg)\\
& = \exp_R\!\bigg(\ii \INT{s}{t}{t'} (\LS(t') - \LS(t')P(t'))\bigg)\\
&\approx \exp_R\!\bigg(\ii \INT{s}{t}{t'} \LS(t')\bigg). 
\end{split}
\end{equation}

Unlike in the usual case, the time-ordered exponential cannot be dropped here. We therefore get
\begin{equation}
\begin{split}
R_{ij}(t,s) &= \INT{0}{1}{\alpha} \Tr\!\bigg(\bar{\rho}(t) e^{\alpha \lambda_l(t)A_l} Q(t)\\
&\qquad\qquad\quad\;\, \exp_R\!\bigg(\ii \INT{s}{t}{t'} \LS(t')\bigg)\ii \LS(t) A_i \\
&\qquad\qquad\quad\;\, e^{-\alpha \lambda_k(t)A_k}\ii \LS(s) A_j\bigg) \\
&\quad\,+ \mathcal{O}((\ii \LS(t) A)^3).
\end{split}\label{rij}\raisetag{12ex}%
\end{equation}
One can then proceed by noticing that the quasi-stationary approximation requires that a clear separation of time scales between slow macroscopic and fast microscopic processes exists. This in turn requires that $t-s \ll \tau_c$ with the characteristic time scale of macroscopic variables $\tau_c$, i.e., the $R_{ij}$ have to decay on a much shorter relaxation time scale $\tau_r \ll \tau_c$. Otherwise, higher-order terms in \cref{taylor} would not be negligible for larger values of $|t-s|$. But if the  $R_{ij}$ vanish for $t-s > \tau_r$, we can extend the time integral in \cref{eom} to $s=-\infty$ without changing the result. This allows in the case of a time-independent Liouvillian, where $G(s,t)$ carries the only $s$ dependence that contributes to \cref{rij}, to write
\begin{equation}
\begin{split}
\INT{0}{t}{s} e^{\ii \LS(t-s)} &\approx \INT{-\infty}{t}{s} e^{\ii \LS(t-s)} 
= - \INT{t}{-\infty}{s} e^{\ii \LS(t-s)} \\
&= \INT{-t}{\infty}{s} e^{\ii \LS(t+s)} 
= \INT{0}{\infty}{s} e^{\ii \LS s}.
\end{split}
\end{equation}
Then, the memory kernel \eqref{AfterEffectFunction} on the right-hand side of Eq.\ \eqref{eom} has effectively no dependence on $t$. Applying this procedure gives in our case
\begin{equation}
\begin{split}
&\INT{0}{t}{s} R_{ij}(t,s) \approx \INT{-\infty }{t}{s} R_{ij}(t,s) = - \INT{t}{-\infty}{s} R_{ij}(t,s) \\
&= \INT{-t}{\infty}{s} R_{ij}(t,-s) = \INT{0}{\infty}{s} R_{ij}(t,t-s).
\end{split}\label{substitutions}%
\end{equation}
Unlike before, one cannot remove the $t$ dependence from the kernel. We therefore get
\begin{equation}
\INT{0}{t}{s} R_{ij}(t,s) \approx D_{ij}(\{a_k(t)\}, t)
\end{equation}
with the diffusion tensor
\begin{equation}
\begin{split}
D_{ij}(\{a_k(t)\}, t) &= \INT{0}{\infty}{s}\!\INT{0}{1}{\alpha} \Tr\!\bigg(\bar{\rho}(\{a_k(t)\}) e^{\alpha \lambda_l(t)A_l} \\
&\quad\,\:\! Q(\{a_k(t)\})\exp_R\!\bigg(\ii \INT{t-s}{t}{t'} \LS(t')\bigg)\ii \LS(t) A_i \\
&\quad\,\:\! e^{-\alpha \lambda_k(t)A_k}\ii \LS(t-s) A_j\bigg).
\end{split}\raisetag{14ex}%
\end{equation}
There is a very notable difference to the usual case discussed by Grabert \cite{Grabert1982}. In that case, the diffusion tensor $D_{ij}$ does not explicitly depend on time, since the time dependence arises only through the $\{a_i(t)\}$ or $\{\mu_i(t)\}$ appearing in the reduced Hamiltonian and the projection operator. In our case, however, even the approximation corresponding to slow variables and time-scale separation still leaves us with an explicitly time-dependent diffusion tensor $D_{ij}(t)$, which will have considerable effects on the phenomenology of transport equations that can be derived from the presented formalism. We can replace $\lambda_j(s)$ by $\lambda_j(t)$ in \cref{eom} as $R_{ij}$ is of second order in $\ii \LS(t)A$ \cite{Grabert1982}. This leads to
\begin{equation}
\dot{a}_i (t) = v_i(t) + D_{ij}(\{a_k(t)\}, t) \lambda_j(t),
\end{equation}
which is the generalized Markovian approximation of the equation of motion \eqref{eom}.

A similar line of argument can be used regarding the time evolution of the fluctuations, assuming that they are also slow. The memory function \eqref{phiij} is of second order in $\ii \LS(t) A$. We can therefore make the quasi-stationary approximation
\begin{equation}%
\begin{split}%
\phi_{ij}(t,s) &= \Tr\!\bigg(\frac{\partial \bar{\rho}(t)}{\partial a_j(t)} \ii \LS(s)Q(t) \\ 
&\qquad\quad\:\! \exp_R\!\bigg(\ii \INT{s}{t}{t'} \LS(t')\bigg) \ii \LS(t) A_i\bigg) \\
&\quad\, - \dot{a}_k(t)\Tr\!\bigg(\frac{\partial^2 \bar{\rho}(t)}{\partial a_j(t) \partial a_k(t)} \\ 
&\qquad\quad\:\! \exp_R\!\bigg(\ii \INT{s}{t}{t'} \LS(t')\bigg)\ii \LS(t) A_i\bigg),
\end{split}\raisetag{15ex}%
\end{equation}%
where we have replaced $a_i(s)$ by $a_i(t)$, $\bar{\rho}(s)$ by $\bar{\rho}(t)$, $P(s)$ by $P(t)$, and $G(s,t)$ by $\exp_R(\ii \TINT{s}{t}{t'} \LS(t'))$. If we now consider the equation of motion for the fluctuations, which is given by \cref{deltaa}, we can replace $\delta A_j(s)$ by $\delta A_j(t)$. This corresponds to the assumption that also the variations of $\delta A_j(t)$ are negligible on the time scale $\tau_r$ on which $\phi_{ij}(t,s)$ decays.\footnote{The memory function $\phi_{ij}$ decays on the same time scale as $R_{ij}$ \cite{Grabert1982}.} The integral over the memory function in \cref{deltaa} can be extended to an integral from $0$ to $\infty$ using the same series of substitutions as in \cref{substitutions}. We can therefore write \cref{deltaa} as
\begin{equation}
\begin{split}
\delta \dot{A}_i(t) &= \Omega_{ij}(t) \delta A_j(t) + D^\mathrm{F}_{ij}(\{a_k(t)\}, t) \delta A_j(t) \\
&\quad\:\! + \delta F_i(t,0)
\end{split}\label{deltaaapprox}%
\end{equation}
with the diffusion tensor for the fluctuations
\begin{equation}
\begin{split}%
D^\mathrm{F}_{ij}(\{a_k(t)\}, t) &= \INT{0}{\infty}{s} 
\Tr\!\bigg(\frac{\partial \bar{\rho}(t)}{\partial a_j(t)} \ii \LS(t-s)Q(t) \\ & \qquad\qquad\;\:\! \exp_R\!\bigg(\ii \INT{t-s}{t}{t'} \LS(t')\bigg) \ii \LS(t) A_i\bigg) \\
&\quad\, - \dot{a}_k(t)\Tr\!\bigg(\frac{\partial^2 \bar{\rho}(t)}{\partial a_j(t) \partial a_k(t)} \\ & \qquad\qquad\;\:\! \exp_R\!\bigg(\ii \INT{t-s}{t}{t'} \LS(t')\bigg)\ii \LS(t) A_i\bigg).
\end{split}\raisetag{12ex}%
\end{equation}

If we are interested in the time evolution of the correlation function \eqref{Cij}, we can employ the same approximation methods as above. The reason is that in the time derivative
\begin{equation}
\frac{\dif}{\dif t} C_{ij}(t,0) =  \bigg(\tdif{}{t} \delta A_i(t), \delta A_j(0)\bigg)
\label{corrsmall}%
\end{equation}
the only $t$ dependence is in the variable $\delta A_i(t)$. Therefore, we can simply insert \cref{deltaaapprox} into \cref{corrsmall} to obtain a differential equation for the correlation function that is correctly expanded in $\ii \LS(t) A$. The result is
\begin{equation}
\frac{\dif}{\dif t} C_{ij}(t,0) = \Omega_{ik}(t) C_{kj}(t,0) + D^\mathrm{F}_{ik}(\{a_k(t)\}, t) C_{kj}(t,0).
\end{equation}

\subsection{\label{bou}Quasi-equilibrium and the Bouchard equation}
To see how our theory connects to previous work, it is important to show that existing results can be derived as a limiting case. Essentially, two theories are relevant here. The first one is the Grabert formalism, which can be applied to systems without time-dependent Hamiltonians far from thermodynamic equilibrium. It has been the basis for many of the derivations above, which it why it is trivial to recover from the discussion above simply by making the assumption that the Liouvillian does not depend on time. The other relevant formalism has been developed by Bouchard as an extension of the Mori theory towards systems with time-dependent Hamiltonians \cite{Bouchard2007}. Being based on the Mori theory, the Bouchard formalism is applicable for small deviations from equilibrium. We can thus derive it as a special case if we linearize our theory around the equilibrium state. However, we first need to discuss what thermal equilibrium implies for the case of time-dependent Hamiltonians, which frequently correspond to external forces driving the system out of equilibrium.

We use a relevant probability density of the canonical form
\begin{equation}
\bar{\rho}(t) = \frac{1}{Z(t)}e^{-\beta (\HS(t) - \mu_j(t)A_j)}.
\label{gc}%
\end{equation}
Here, we have introduced the thermodynamic beta $\beta=1/(k_B T)$ with the Boltzmann constant $k_B$ and temperature $T$ as well as the conjugate forces $\{\mu_i(t)\}$, which work like the $\{\lambda_i(t)\}$ used in \cref{derivation} (i.e., they ensure $\Tr(\bar{\rho}(t) A_i) = a_i(t)$). The advantage of the canonical form is that we have separated the Hamiltonian, which governs the time evolution, from the rest of the operators, which will be useful for re-writing the equations in the way sketched below.\footnote{In this case, one needs to find an appropriate definition of the temperature $T$, e.g., by referring to a reservoir temperature or to equilibrium states corresponding to mean values. For a discussion, see Ref.\ \cite{Grabert1982}.} 
If one introduces the coarse-grained Helmholtz free energy
\begin{equation}
\mathcal{F}(t) = \Tr(\bar{\rho}(t)\HS(t)) + k_B T \Tr(\bar{\rho}(t)\ln(\bar{\rho}(t))),
\end{equation}
the conjugate forces are given by \cite{Grabert1982}
\begin{equation}
\mu_i(t) = \frac{\partial \mathcal{F}(t)}{\partial a_i(t)}.
\end{equation}
The equation of motion \eqref{eom} can be written as
\begin{equation}
\dot{a}_i(t) = - V_{ij}(t) \mu_j(t)  - \INT{0}{t}{s} R_{ij}(t,s) \beta\mu_j(s) 
\label{eomRij}%
\end{equation}
with the drift tensor
\begin{equation}
V_{ij}(t) = \Tr\!\bigg(\bar{\rho}(t) \frac{\ii}{\hbar} [A_i, A_j]\bigg).
\label{Vij}%
\end{equation}
This results from 
\begin{equation}
\begin{split}
v_i(t) &= \Tr(\bar{\rho}(t)\ii \LS(t)A_i)
= - \Tr((\ii \LS(t) \bar{\rho}(t))A_i) \\
&= - \Tr\!\bigg(\frac{\ii}{\hbar}[ \HS(t), \bar{\rho}(t)]A_i\bigg)\\
&= - \Tr\!\bigg(\frac{\ii}{\hbar}\bigg[ \HS(t) - \mu_j(t) A_j + \mu_j(t)A_j,\\ 
&\qquad\qquad \frac{1}{Z(t)}e^{-\beta(\HS(t) - \mu_k(t) A_k)}\bigg]A_i\bigg)\\
&= - \Tr\!\bigg(\frac{\ii}{\hbar}\bigg[ \mu_j(t)A_j, \frac{1}{Z(t)}e^{-\beta(\HS(t) - \mu_k(t) A_k)}\bigg]A_i\bigg)\\
&= - \Tr\!\bigg(\frac{\ii}{\hbar}\frac{1}{Z(t)}e^{-\beta(\HS(t) - \mu_k(t) A_k)}[\mu_j(t)A_i, A_j]\bigg)\\
&= - \Tr\!\bigg(\bar{\rho}(t) \frac{\ii}{\hbar}[A_i, A_j]\bigg) \mu_j(t).
\end{split}\label{vit}\raisetag{23ex}%
\end{equation}
Due to the change to the canonical relevant probability density \eqref{gc}, the microcanonical form \eqref{RetardationMatrix} of the retardation matrix has to be modified here by inserting the canonical relevant density \eqref{gc} for $\bar{\rho}(s)$ and replacing $\lambda_j(t) A_j \to \beta (\HS(t)-\mu_j(t)A_j)$.

We now assume that all thermodynamic forces vanish, i.e., $\mu_i(t) = 0$. In the canonical ensemble, the probability density of a system with a time-independent Hamiltonian $H$ in thermodynamic equilibrium is
\begin{equation}
\hat{\rho} = \frac{1}{Z}e^{-\beta H},
\end{equation}
where, similar to $Z(t)$ in \cref{rhobar,gc}, the partition function $Z$ is the normalization factor. This state is constant in time. There are two possibilities for getting a similar situation in the case of a time dependence of the Hamiltonian:
\begin{enumerate}
\item The Hamiltonian is given by $\HS(t) = \HSO + \delta \HS(t)$, where $\HSO$ is time-independent and $\delta \HS(t)$ is a small perturbation. Then we can, to a good approximation, write
\begin{equation}
\bar{\rho}(t) = \frac{1}{Z(t)}e^{-\beta \HS(t)} \approx \frac{1}{Z}e^{-\beta \HSO} = \hat{\rho}
\end{equation}
and use $\hat{\rho}$ to evaluate, e.g., correlation functions. 
\item The Hamiltonian $\HS(t)$ is varying very slowly in time, such that it is reasonable to assume that the system always has enough time to adjust to the current value of $\HS(t)$ and to achieve a quasi-equilibrium with respect to this value. One could call this an adiabatic approximation.
\end{enumerate}

For the calculations in this section, we consider the first scenario, in which the time-dependent part of the Hamiltonian is small compared to the time-independent part. The reason is that in this way we can recover Bouchard's results, which suggests that this scenario corresponds to the limit in which they can be applied. Moreover, this allows to avoid the complications associated with explicitly time-dependent relevant densities. However, it is also possible to make a quasi-equilibrium approximation based on the second scenario in a similar fashion. We assume that $\beta\delta \HS(t)$ is small, which is valid for a sufficiently small $\beta$, but not that $\LS(t) = \LSO + \delta \LS(t) \approx \LSO$, because this approximation would only recover the standard Mori theory. Our assumption is justified, because we formally want our equation of motion \eqref{eomRij} to be of linear order in the small quantities $\beta\delta \HS(t)$ and $\beta \mu_j(t) A_j$. Therefore, when $\mu_j(t) \neq 0$ holds and we assume that also $\mu_j(t)$ itself is small, all contributions of $\beta\delta \HS(t)$ to the tensors $V_{ij}$ and $R_{ij}$ have to be dropped and Mori products are evaluated using $\hat{\rho}$. However, if temperatures are sufficiently large, it is possible that the time-dependent part $\delta \LS(t) = \frac{1}{\hbar} [\delta \HS(t), \cdot]$ of the Liouvillian is not negligible even though $\beta\delta \HS(t)$ is.

The first assumption that is required is that the variables $\{A_i\}$ are defined in such a way that they vanish in thermodynamic equilibrium, i.e.,
\begin{equation}
\langle A_i \rangle_{\mathrm{eq}} = 0.
\label{null}%
\end{equation}
In the following, we use $\langle \cdot \rangle_{\mathrm{eq}}=\Tr(\hat{\rho}\,\cdot)$ to indicate that the operators are evaluated at equilibrium. Equation \eqref{null} is not a problematic restriction here and in fact common in classical Mori theory \cite{HolianE1985}, in particular because one can re-define the variables in such a way that this is true. However, it might be more problematic if one considers the second, i.e., adiabatic, scenario, because then it is possible that the mean values change over time, although slowly. In this scenario one would either have to make the definition of the operator time-dependent, or the assumption \eqref{null} has to be dropped leading to additional terms in the resulting equations. From \cref{null} we get
\begin{equation}
\delta A \approx A - \langle A \rangle_{\mathrm{eq}} = A
\end{equation}
so that we can simply write $A$ for $\delta A$.

A further simplification concerns the form of the correlator. It is notable that in the equations relevant for this calculation, correlations are always taken at equal times (i.e., we calculate $(X(t), Y(t))$, but not $(X(t), Y(s))$ for $t \neq s$). In the usual Mori theory, one would simply write \cite{Grabert1982}
\begin{equation}
(X, Y)_M = \INT{0}{1}{\alpha} \Tr\!\big(\bar{\rho}(t)Xe^{-\alpha \beta H} Y^\dagger e^{\alpha \beta H}\big)
\label{moriproduct}%
\end{equation}
and take this to be the scalar product. For recovering this Mori product in the situation close to thermodynamic equilibrium from our correlation function, take the general form \eqref{correlation} with $s=t$, replace $\lambda_j(t) A_j \to \beta \HSO$, which is possible in thermal equilibrium when transiting from the microcanonical relevant density \eqref{rhobar} to the canonical one \eqref{gc}, and set generally $\exp_L(- \ii\TINT{0}{t}{t'} \LS(t')) X(t) = X$. This works even in the adiabatic scenario.\footnote{The approximation $\HS(t) \approx \HSO$ ensures that the projection operator constructed from the Mori product \eqref{moriproduct} is time-independent, which is assumed in Bouchard's derivation. Dropping this assumption, which is necessary in the adiabatic scenario, leads to more complex and more general equations.}

These assumptions can be used to re-write \cref{projectioncorr} for the projection operator. If we use $\bar{\rho}(t) = \hat{\rho}$ and assume that our observables vanish at equilibrium, we can write the normalization matrix as
\begin{equation}
(\delta A_i(t), \delta A_j(t)) \approx (A_i, A_j)_M.
\label{morip}%
\end{equation}
The first term on the right-hand side of \cref{projectioncorr} vanishes, since it corresponds to the value at equilibrium. From the second term, one gets 
\begin{equation}
PX = A_j (A_j, A_k)_M^{-1} (X, A_k)_M,
\label{prolin}%
\end{equation}
which is the Mori projection operator \cite{Mori1965, Grabert1982}. Notably, it is time-independent, unlike the general projection operator \eqref{projectioncorr} \cite{Grabert1982}.

Next, we consider small deviations from equilibrium, where $a(t)$ does not vanish. We therefore use the canonical form \eqref{gc} for the relevant density $\bar{\rho}(t)$ and assume that our thermodynamic forces $\{\mu_i(t)\}$ are small. This gives\footnote{We could in principle also perform the expansion in $\beta \delta \HS(t)$, but this would not be practically useful.} 
\begin{equation}
\begin{split}
\bar{\rho}(t) &= \frac{1}{Z(t)}e^{-\beta (\HS(t) -\mu_j(t) A_j)}\\ 
&\approx \frac{1}{Z(t)}e^{-\beta \HS(t)} e^{\beta \mu_j(t) A_j}e^{\frac{\beta^2}{2}[ \HS(t),\mu_k(t) A_k]}\\
&\approx \frac{1}{Z(t)}e^{-\beta \HS(t)} \bigg( 1 + \beta \mu_j(t) A_j + \frac{\beta^2}{2}[ \HS(t),\mu_k(t) A_k]\bigg)\\
&\approx \frac{1}{Z(t)}e^{-\beta \HS(t)} (1 + \beta \mu_j(t) A_j),
\end{split}\label{approx}\raisetag{15ex}%
\end{equation}
where we have used the Baker-Campbell-Hausdorff formula \cite{BlanesCOR2010}
\begin{equation}
\exp(X)\exp(Y) = \exp\!\Big(X + Y + \frac{1}{2}[X,Y] + \dotsb\Big).
\label{bch}%
\end{equation}
The expansion in \cref{approx} is carried out to first order in $\beta$ and $\mu_i(t)$. For close-to-equilibrium systems at very low temperatures, where $\beta$ is not small, one has to use
\begin{equation}
\begin{split}
&\exp(\beta \mu_j(t) A_j)\exp(-\beta \HS(t)) \\
&= \exp\!\bigg(\! -\beta \HS(t) + \sum_{n=0}^{\infty}(-1)^n\beta^{n+1} \mu_j(t)\frac{B_n}{n!}[\HS(t),A_j]_n \\
&\qquad\quad\;\, + \mathcal{O}(\mu_i(t)^2)\bigg).
\end{split}\raisetag{8ex}%
\end{equation}
This equation is based on a simplified form of the Baker-Campbell-Hausdorff formula, where $[X,Y]_n = [X,\dotsb [X,Y]\dotsb]$ with $[X,Y]_0 = Y$ is an $n$-times nested commutator and $B_n$ is a Bernoulli number with $B_1=-1/2$ \cite{BlanesCOR2010}. For a discussion of the convergence of the expansion \eqref{bch}, see Ref.\ \cite{BlanesCOR2009}.

The equation \cite{Grabert1982}
\begin{equation}
\frac{\partial a_i(t)}{\partial \mu_j(t)} = \beta (\delta A_i(t), \delta A_j(t))
\end{equation}
gives, using \cref{morip}, to linear order in deviations from equilibrium
\begin{equation}
\mu_i(t) = \frac{a_j(t)}{\beta}(A_j, A_i)_M^{-1}.
\end{equation}
Therefore, with \cref{approx} we obtain
\begin{equation}
\bar{\rho}(t) \approx \frac{1}{Z(t)}e^{-\beta \HS(t)} \big(1 + a_k(t)A_j(A_k, A_j)_M^{-1}\big).
\label{lin}%
\end{equation}
It is important that \cref{lin} is linear in $a_i(t)$. Our evolution equation \eqref{eomRij} for $a_i(t)$ becomes
\begin{equation}
\dot{a}_i(t) = \Omega_{ij}(t) a_j(t) + \INT{0}{t}{s} \mathcal{H}_{ij}(t,s) a_j(s),
\label{bouchardmean}%
\end{equation}
where we have defined the frequency matrix
\begin{equation}
\Omega_{ij}(t) = - \frac{1}{\beta} V_{ik}(t) (A_j, A_k)_M^{-1}
\label{firstdefinition}%
\end{equation}
and the memory matrix
\begin{equation}
\mathcal{H}_{ij}(t,s) = -R_{ik}(t,s) (A_j, A_k)_M^{-1}.
\label{Hijts}%
\end{equation}
Moreover, we can consider our general evolution equation \eqref{deltaa} for deviations from the mean values. It is here
\begin{equation}
\delta \dot{A}_i(t) = \Omega_{ij}(t) \delta A_j(t) + \INT{0}{t}{s} \phi_{ij}(t,s) \delta A_j(s) + F_i(t,0).
\end{equation}
The frequency matrix can also be written as
\begin{equation}
\Omega_{ij}(t) = (A_j, A_k)_M^{-1} (\ii \LS(t)A_i, A_k)_M.
\label{seconddefinition}%
\end{equation}
Equation \eqref{seconddefinition} is identical with Bouchard's definition \cite{Bouchard2007}. To see that \cref{firstdefinition,seconddefinition} are equivalent, note that
\begin{equation}
\begin{split}
V_{ij}(t) &= - \frac{\partial v_i(t)}{\partial \mu_j(t)}= \beta \frac{\partial v_i(t)}{\partial \lambda_j(t)} \\
&= - \beta(\dot{A}_i(t), \delta A_j(t))\\
&\approx -\beta (\ii \LS(t) A_i, A_j)_M,
\end{split}
\end{equation}
where we have used Eqs.\ \eqref{Vij}, \eqref{vit}, and \eqref{partialvl} (see Appendix \ref{appendixa}). 
For calculating $\phi_{ij}$, we can use the fact that \cite{Grabert1982}\footnote{The relation $\phi_{ij}(t,s) = \delta w_i(t)/\delta a_j(s)$ can be verified by explicitly calculating the functional derivative of Eq.\ \eqref{DisorganizedDrift} using Eq.\ \eqref{average} in the general case \cite{Grabert1982}.}
\begin{equation}
\begin{split}
\phi_{ij}(t,s) &= \frac{\delta \ww_i(t)}{\delta a_j(s)} = \frac{\delta}{\delta a_j(s)} \INT{0}{t}{t'} K_i(t,t')\\
& = - \frac{\delta}{\delta a_j(s)} \INT{0}{t}{t'} R_{ik}(t,t')a_l(t')(A_l, A_k)_M^{-1}.
\end{split}\label{fd}\raisetag{8ex}%
\end{equation}
This follows from Eqs.\ \eqref{DisorganizedDrift}, \eqref{average}, \eqref{bouchardmean}, and \eqref{Hijts}.
At equilibrium, we have
\begin{equation}
\begin{split}
R_{ij}(t,s) &= \INT{0}{1}{\alpha} \Tr\!\big(\hat{\rho}\, e^{\alpha \beta H_0} Q \\
&\qquad\qquad\quad\;G(s,t)\ii \LS(t) A_i e^{-\alpha \beta H_0}\ii \LS(s) A_j\big).
\end{split}\raisetag{3em}%
\end{equation}
The relevant density does not depend on $a_i(t)$ at equilibrium. For the projection operator, the $a_i(t)$ dependence is \cite{Grabert1982}
\begin{equation}
\frac{\partial}{\partial a_i(t)}P(t) X = (A_j - a_j(t)) 
\Tr\!\bigg(\frac{\partial^2 \bar{\rho}}{\partial a_i(t) \partial a_j(t)}X\bigg).
\end{equation}
This vanishes as $\bar{\rho}$ is linear in $a_i$ in the regime we consider. Therefore, we can immediately evaluate the functional derivative in \cref{fd} and get\footnote{We write $Q$ instead of $Q(s)$, since $Q$ is time-independent in this approximation.}
\begin{equation}
\begin{split}
&\,\phi_{ij}(t,s) \\
&= - R_{ik}(t,s) (A_j, A_k)_M^{-1} = \mathcal{H}_{ij}(t,s)\\
&= - (A_j, A_k)_M^{-1} \INT{0}{1}{\alpha} \Tr\!\big(\hat{\rho}\, e^{\alpha\beta H_0}QG(s,t)\ii \LS(t) A_i\\
&\qquad\qquad\qquad\qquad\qquad\;\;\; e^{-\alpha\beta H_0}\ii \LS(s) A_k\big)\\
&= - (A_j, A_k)_M^{-1} \big(QG(s,t)\ii \LS(t) A_i, \ii \LS(s) A_k\big)_M.
\end{split}\raisetag{15ex}%
\end{equation}
Using the fact that the Mori product is antisymmetric with respect to the Liouvillian,\footnote{Equation \eqref{Mori_product_antisymmetric} holds also when using the Liouvillian $\LH$ instead of $\LS$.} i.e., \cite{Grabert1982}
\begin{equation}
(X, \ii \LS Y)_M = - (\ii \LS X, Y)_M,
\label{Mori_product_antisymmetric}%
\end{equation}
we get
\begin{equation}
\phi_{ij}(t,s) = (A_j, A_k)_M^{-1} \big(\ii \LS(s) Q G(s,t)\ii \LS(t) A_i, A_k\big)_M.
\label{fiij}%
\end{equation}
With the definition of the random force
\begin{equation}
F_i(t,s) = QG(s,t)\ii \LS(t) A_i = G(s,t)Q\ii \LS(t) A_i,
\end{equation}
we can write \cref{fiij} as
\begin{equation}
\phi_{ij}(t,s) = \mathcal{H}_{ij}(t,s) = (A_j, A_k)_M^{-1} (\ii \LS(s) F_i(t,s), A_k)_M
\label{phiijBouchard}%
\end{equation}
and our evolution equation is
\begin{equation}
\begin{split}
\delta \dot{A}_i(t) &= \Omega_{ij}(t) \delta A_j(t) + \INT{0}{t}{s} \mathcal{H}_{ij}(t,s) \delta A_j(s) \\ 
&\quad\:\! + F_i(t,0).
\end{split}\label{deltabouchard}%
\end{equation}

Comparing \cref{bouchardmean,deltabouchard} shows that the mean values $\{a_i(t)\}$ and the fluctuations $\{\delta A_i(t) \}$ follow an analogous evolution law in the vicinity of thermodynamic equilibrium. Adding \cref{bouchardmean,deltabouchard} gives
\begin{equation}
\begin{split}
\dot{A}_i(t) &= \Omega_{ij}(t) A_j(t) + \INT{0}{t}{s} \mathcal{H}_{ij}(t,s) A_j(s) \\ 
&\quad\:\! + F_i(t,0).
\end{split}\label{bouchard}%
\end{equation}
This is identical with Bouchard's result, which he derives by applying the identity \eqref{dyson} to $Q\ii \LS(t)A_i$. As in \cref{TimeEvolutionCij}, we can easily derive an equation of motion for the correlation functions:
\begin{equation}
\dot{C}_{ij}(t,0) = \Omega_{ik}(t)C_{kj}(t,0) + \INT{0}{t}{s} \mathcal{H}_{ik}(t,s)C_{kj}(s,0).
\label{simplecorrelation}%
\end{equation}

There is a conceptually interesting point one can make about the form of \cref{bouchard}. Bouchard compares it with the usual Mori equation \cite{Bouchard2007}
\begin{equation}
\dot{A}_i(t) = \Omega_{ij} A_j (t) + \INT{0}{t}{s} \mathcal{H}_{ij}(s) A_j(t-s) + F_i(t),
\label{mori}%
\end{equation}
where
\begin{equation}
\mathcal{H}_{ij}(s) = (A_j, A_k)_M^{-1} (\ii L F_i(s),A_k)_M
\end{equation}
and
\begin{equation}
F_i(t) = e^{Q\ii Lt} Q \ii LA_i,
\end{equation}
which is a special case of the Bouchard result for a time-independent Liouvillian $L$. He mentions that a major difference between his result and the Mori equation is the fact that his memory kernel depends on $t$ and $s$ rather than $t-s$, which he correctly attributes to the time dependence of the Hamiltonian \cite{Bouchard2007}. Interestingly, the same is already true for the Grabert formalism, in which the Liouvillian is time-independent. 
The reason for this is the factor $G(s,t)$, which is a time-ordered exponential involving the Liouvillian and the projection operator. In the Mori formalism, both $L$ and $Q$ are time-independent. This is why one can re-write $G(s,t)$ as a normal exponential. Moreover, the propagator $\exp_R(\ii\TINT{0}{s}{t'} \LS(t'))$ simplifies, after the substitution $s\to t-s$, to $e^{\ii L(t-s)}$, leading to a dependence only on time differences.
For this to be no longer possible, it is sufficient that the Liouvillian or the projection operator is time-dependent. In the far-from equilibrium case treated by Grabert, the projection operator is time-dependent and the Liouvillian is not, whereas Bouchard has a time-dependent Liouvillian combined with a time-independent projection operator. In the formalism presented in this work, both operators are time-dependent. All three scenarios lead to a time-ordered exponential $G(s,t)$ and therefore to a memory kernel that depends on distinct times rather than on a time difference. 

The linearized form of the Mori-Zwanzig theory has an interesting connection to linear irreversible thermodynamics \cite{BrandPB2018}. This theory allows, on a phenomenological level, to describe systems that are in local thermodynamic equilibrium. It is closely related to the usual Mori-Zwanzig theory for time-independent Hamiltonians, which allows, if linearized around thermodynamic equilibrium \cite{Grabert1982}, to derive important relations of linear irreversible thermodynamics, such as the Onsager reciprocal relations and the fact that entropy does not decrease \cite{Forster1989}. These calculations are based on the properties of the relevant observables under time-reversal, which is even or odd \cite{Grabert1982, Forster1989}. In the case of small deviations from thermodynamic equilibrium in the presence of a time-dependent Hamiltonian, which is considered in this work, the situation is more subtle. It is possible that the time dependence of the Hamiltonian breaks the microscopic time-reversal invariance. In this case, the systems can violate the Onsager reciprocal relations. Comparing macroscopic equations derived using \cref{bouchard} with the usual results of linear irreversible thermodynamics might therefore give deeper insights into the influence of time-dependent Hamiltonians on the thermodynamic properties of a system, e.g., regarding symmetries or irreversibility.

Finally, the range of validity for \cref{bouchard} is, as well, a difficult question: Bouchard derives it by applying the generalized Dyson identity \eqref{dyson} to $Q\ii \LS(t)A_i$, where the projection operator $P$ is given by \cref{prolin} and $(X,Y)$ is a scalar product, which he does not specify, although he discusses various possible choices including the Mori product \cite{Bouchard2007}. Since this derivation does not assume that the system is close to equilibrium or at sufficiently high temperatures, his result is always formally correct. In the far-from-equilibrium case, his definition of the projection operator differs from the definition \eqref{projection} used in this work. The problem is that time-independent projection operators cannot appropriately describe the far-from-equilibrium dynamics, which is typically nonlinear \cite{Grabert1982}. Since a time-independent projection operator leads to a linear Langevin equation, nonlinear contributions will be projected out together with the irrelevant noise \cite{Bouchard2007}, so that the resulting equations provide fewer insights into the relevant dynamics. Therefore, the derivation presented in this section allows for a better understanding of the conditions under which the Bouchard equation should be applied.

\subsection{Expansions for the time-ordered exponentials}
Although the equations of motion resulting from our theory are formally exact, the time-ordered exponentials might be difficult to evaluate in practice. A common tool for doing this is the Magnus expansion \cite{Magnus1954, BlanesCOR2010}. Given an equation
\begin{equation}
\tdif{}{t} A(t) = \ii \LS(t)A(t),
\end{equation}
it allows to compute an operator $M(t,0)$ such that
\begin{equation}
A(t) = e^{M(t,0)}A,
\end{equation}
which is a true exponential solution that can be used to define an effective Liouvillian $L^{\mathrm{eff}}(t,0)=-\ii M(t,0)/t$. This can then be used to replace the time-ordered exponentials. Although no closed equation for $M(t,0)$ exists, the Baker-Campbell-Hausdorff formula \eqref{bch} allows for a series expansion of $M(t,0)$ in terms of commutators of the Liouvillians. This is particularly useful, because a variety of numerical methods exist that allow to calculate these terms to any finite order \cite{BlanesCOR2009}. Since this is a standard derivation\footnote{One should, however, be careful, because the typical discussions are based on left-time-order, whereas here a right-time-ordered exponential is used.} that can be found in a number of sources \cite{OteoR2000, BlanesCOR2009, BlanesCOR2010}, we do not present it here and instead discuss some aspects that are important for the problem at hand.

The Magnus expansion is based on an expansion in commutators of the Liouvillians at different points in time. The convergence of a Magnus expansion for a time-ordered exponential depends both on $\norm{\LS}$, where $\norm{\cdot}$ denotes the operator norm, and on the range of the time interval that is integrated over (it should be small). Hence, the Magnus expansion is particularly useful for situations in which the observables of interest vary slowly so that $\ii \LS(t) A$ is small. Therefore, it will typically be combined with the generalized Markovian approximation. In this case, it is used for the time-ordered exponential $\exp_R(\ii \TINT{t-s}{t}{t'} \LS(t'))$, which results from applying the Markovian approximation and the substitutions from \cref{substitutions} to $G(s,t)$. Although this exponential is integrated over from $s=0$ to $s=\infty$, it is a part of the memory kernel that is assumed to vanish for larger values of $t-s$, which means that one can restrict the calculation to a certain interval. The precise convergence conditions of the Magnus expansion depend on the mathematical properties of the operator whose exponential is expanded \cite{OteoR2000}. Moreover, the important case of periodic time dependence is particularly good to handle, since in those cases symmetry arguments allow to cancel certain terms in the expansion \cite{ApperleyHH2012}.

If the Hamiltonian in question is a sum of a time-independent part and a time-dependent part, we can write the Liouvillian as
\begin{equation}
\LS(t) = \LSO + \LSI(t).
\end{equation}
The typical approach to this type of equations is to treat $\LSI(t)$ as a small perturbation: Since we need the Magnus expansion only for $\LSI$, the relevant parameter $\norm{\LS}$ will be smaller and we can use the expansion on much longer time scales. This means that we need a way of separating the time evolutions. If $[\LSO, \LSI(t)] = 0$, we can simply write
\begin{equation}
\begin{split}
\exp_R\!\bigg(\ii \INT{0}{t}{t'} \LS(t')\bigg) &= \exp_R\!\bigg(\ii \LSO t + \ii  \INT{0}{t}{t'} \LSI(t')\bigg) \\
&= e^{\ii \LSO t} \exp_R\!\bigg(\ii \INT{0}{t}{t'}\LSI(t')\bigg)
\end{split}\raisetag{7ex}%
\end{equation}
and perform the expansion for the second factor only, giving the result
\begin{equation}
A(t) = e^{\ii (\LSO + L^{\mathrm{eff}}_1(t,0))t} A,
\end{equation}
where $L^{\mathrm{eff}}_1 (t,0) =-\ii M(t,0)/t$ is the effective Liouvillian. This result can then be used as if the Hamiltonian were time-independent. Unfortunately, this is no longer possible, if the two parts of the Liouvillian do not commute, since then $e^X e^Y \neq e^{X+Y}$. In this case, we can define the modified operators
\begin{equation}
A_I(t) = e^{- \ii \LSO t} A(t)
\end{equation}
and
\begin{equation}
\LLI(t) = e^{-\ii \LSO t} \LSI(t) e^{\ii \LSO t},
\end{equation}
where the subscript \ZT{I} is motivated by the formal analogy of this procedure to the usual interaction picture \cite{PeskinS1995}. Calculating the time evolution of $A_I(t)$ gives
\begin{equation}
\begin{split}
\tdif{}{t} A_I(t) &= \tdif{}{t}e^{-\ii \LSO t}A(t)\\
&= -\,e^{-\ii \LSO t} \ii \LSO A(t) + e^{- \ii \LSO t} \ii \LSO A(t) \\ 
&\quad\:\! + e^{- \ii \LSO t} \ii \LSI(t) A(t) \\
&= e^{- \ii \LSO t} \ii \LSI(t) e^{\ii \LSO t}e^{- \ii \LSO t} A(t)\\ 
&= \ii \LLI (t) A_I(t).
\end{split}\raisetag{10ex}%
\end{equation}
The equation for $A_I(t)$ can then be solved, giving
\begin{equation}
A_I(t) = \exp_L\!\bigg(\ii \INT{t_0}{t}{t'} \LLI(t')\bigg) A_I(t_0).
\end{equation}
Transforming back then shows
\begin{equation}
\begin{split}
A(t) &= e^{\ii \LSO t} A_I(t) \\
&= e^{\ii \LSO t} \exp_L\!\bigg(\ii \INT{t_0}{t}{t'} \LLI(t')\bigg) e^{-\ii \LSO t_0} A(t_0).
\end{split}\label{backtransform}\raisetag{8ex}%
\end{equation}
For the right-time-ordered exponential, we use the fact that 
\begin{equation}
A(t) = \exp_R\!\bigg(\ii \INT{t_0}{t}{t'} \LS(t')\bigg) A(t_0)
\end{equation}
is the formal solution of $\dot{A}(t) = -\ii \LS(t)A(t)$ for a $t < t_0$. By the argument that has lead to \cref{backtransform}, that equation is also solved by
\begin{equation}
A(t) =  e^{-\ii \LSO t} \exp_R\!\bigg(\ii \INT{t_0}{t}{t'} \LRI(t')\bigg) e^{\ii \LSO t_0} A(t_0),
\end{equation}
where
\begin{equation}
\LRI(t) = e^{\ii \LSO t} \LSI(t) e^{-\ii \LSO t}.
\end{equation}
Therefore, we have
\begin{equation}
\begin{split}
&\exp_R\!\bigg(\ii \INT{t_0}{t}{t'} \LS (t')\bigg) \\
&\!\:\!= e^{-\ii \LSO t} \exp_R\!\bigg(\ii \INT{t_0}{t}{t'} \LRI(t')\bigg) e^{\ii \LSO t_0}.
\end{split}
\end{equation}
Replacing $t_0$ by $t-s$, this shows that
\begin{equation}
\begin{split}
&\exp_R\!\bigg(\ii \INT{t-s}{t}{t'} \LS(t')\bigg) \\
&\!\:\!= e^{-\ii \LSO t} \exp_R\!\bigg(\ii \INT{t-s}{t}{t'} \LRI(t')\bigg) e^{\ii \LSO (t-s)}.
\end{split}
\end{equation}
One can write
\begin{equation}
\exp_R\!\bigg(\ii \INT{t-s}{t}{t'} \LRI(t')\bigg) = e^{\ii L^{\mathrm{eff}}_I(t,t-s) s},
\end{equation}
where $L^{\mathrm{eff}}_I(t,t-s)$ is determined by the Magnus expansion, which has much better convergence for $\LRI$ than for $\LS$. This gives
\begin{equation}
\exp_R\!\bigg(\ii \INT{t-s}{t}{t'} \LS(t')\bigg) = e^{-\ii \LSO t} e^{\ii L^{\mathrm{eff}}_I(t,t-s) s} e^{\ii \LSO (t-s)}.
\end{equation}
Note that we can write this as 
\begin{equation}
e^{\ii  (-\LSO t + L^{\mathrm{eff}}_I(t,t-s) s + \LSO(t-s))} = e^{\ii (L^{\mathrm{eff}}_I(t,t-s) - \LSO) s}
\end{equation}
only if $[\LSO, L^{\mathrm{eff}}_I] = 0$, which is not generally the case. 

Another interesting modification concerns \textit{time-convolutionless equations}. For various types of projection operator methods, these can be applied to re-write the equations in such a way that no time-convolutions appear. In the case of the Mori-Zwanzig formalism, these appear through the memory kernels, as can be seen particularly clearly from \cref{mori}, and are usually eliminated, e.g., through the Markovian approximation, which makes the resulting equations easier to handle. Time-convolutionless projection operators form an alternative. In the case of time-independent Hamiltonians, time-convolutionless projection operator methods for statistical mechanics in the form of the Mori-Zwanzig formalism already exist \cite{ChaturvediS1979}. For other projection operator methods, convolutionless equations have also been derived in the case of time-dependent Hamiltonians \cite{KoideM2000}. These are used, e.g., in quantum field theory \cite{KoideMT1999}. Combining those approaches with our formalism could also help to develop further approximation methods. However, a detailed treatment of this is beyond the scope of this work.

\subsection{Classical limit}
While the previous treatment has been quantum mechanical in order to keep it as general as possible, the formalism can also be applied to classical systems. All of the previous results still hold in this limiting case. However, there are three changes, which simplify the equations:
\begin{enumerate}
\item The Hilbert space operators $\{A_i\}$ are now functions $\{A_i(\vec{q}(t),\vec{p}(t))\}$ defined on the phase space, where $\vec{q}$ and $\vec{p}$ are vectors containing the canonical coordinates and momenta of all particles, respectively.
What is central to our calculation is the correct definition of the Schr\"odinger picture Liouvillian in this case. As Holian and Evans have shown \cite{HolianE1985,EvansM2008}, the time evolution of a function defined on the phase space is given by
\begin{equation}
A(\vec{q}(t),\vec{p}(t)) = \exp_R\!\bigg(\ii \INT{0}{t}{t'} L_p(t')\bigg) A(\vec{q}(0),\vec{p}(0)),
\end{equation}
where they introduce the $p$ Liouvillian defined as\footnote{In Refs.\ \cite{HolianE1985,EvansM2008}, the Poisson bracket notation is not used, but the definitions appearing there can be written using Poisson brackets.}
\begin{equation}
\ii L_p(t) = \{H(\vec{q},\vec{p},t), \cdot \}.
\end{equation}
Here, $\{\cdot,\cdot\}$ denotes the Poisson bracket. The $p$ Liouvillian depends on the current Hamiltonian, but acts on the initial phase state. Therefore, it agrees with our Schr\"odinger picture Liouvillian.\footnote{Note that our approach is slightly different from the treatment by Holian and Evans. In Refs.\ \cite{HolianE1985,EvansM2008}, the $p$ Liouvillian is introduced for the Heisenberg picture, i.e., they directly define the Liouvillian in such a way that it acts in right time order on the observables. The $f$ Liouvillian, which they also introduce, is not identical with the Heisenberg picture Liouvillian used here. Instead, it corresponds to the Schr\"odinger picture Liouvillian appearing in \cref{vonneu}.}
\item The trace Tr corresponds to a phase space integral rather than to the quantum-mechanical trace:
\begin{equation}
\Tr(X(t)) = \frac{1}{h^{3N} N!}\INT{}{}{^{3N}q} \!\INT{}{}{^{3N}p} X(\vec{q},\vec{p},t).
\end{equation}
Here, $N$ is the number of particles and $h$ is the Planck constant.
\item Since functions, unlike operators, generally commute, the correlator \eqref{correlation} can be simplified to 
\begin{equation}
\begin{split}
(X(t), Y(s)) \!\!\!&\\
= \Tr\!\bigg(\bar{\rho}(s) \,& \bigg(\exp_L\!\bigg(\!- \ii\INT{0}{s}{t'} \LS(t')\bigg) X(t)\bigg)\\
&\bigg(\exp_L\!\bigg(\!- \ii\INT{0}{s}{t'} \LS(t')\bigg) Y(s) \bigg)\!\bigg).
\end{split}\label{CorrelationClassical}\raisetag{8ex}%
\end{equation}
For systems that are at time $s$ in a state described by the relevant density, i.e., for $\rho(s)=\bar{\rho}(s)$, \cref{CorrelationClassical} reduces to
\begin{equation}
(X(t), Y(s)) = \Tr(\rho(0)X(t)Y(s)).
\label{CorrelationClassicalSimple}%
\end{equation}
This is the correct definition of the classical correlation function \cite{Grabert1982}.
The reason for obtaining \cref{CorrelationClassicalSimple} is that $\Tr(\rho(s)XY)=\Tr((\exp_L(- \ii \TINT{0}{s}{t'} \LS(t'))\rho(0))XY)$. As described in \cref{time}, the latter expression can be re-written as $\Tr(\rho(0)(\exp_R( \ii \TINT{0}{s}{t'} \LS(t')) (XY)))$. Applying this together with the identity
\begin{equation}
\begin{split}
\exp_R\!\bigg(\ii \INT{0}{t}{t'} \LS(t')\bigg)(XY) &= \bigg(\!\exp_R\!\bigg(\ii \INT{0}{t}{t'} \LS(t')\bigg) X\bigg) \\
&\quad\;\:\! \bigg(\!\exp_R\!\bigg(\ii \INT{0}{t}{t'} \LS(t')\bigg) Y\bigg)
\end{split}\label{expRrelation}\raisetag{7.5ex}%
\end{equation}
to \cref{CorrelationClassical} immediately leads to \cref{CorrelationClassicalSimple}. Equation \eqref{expRrelation} can be proven by writing out the exponentials as a series, sorting the terms by orders in $\LS$, and using the product rule.
\end{enumerate}

\section{\label{spins}Application to spin relaxation}
To demonstrate the usefulness of our new theory, we consider spin relaxation in external magnetic fields. This is a very interesting case, since it is important for NMR, which has many applications ranging from chemistry \cite{CobbM2008} to medicine \cite{YoungHBLS1981}. NMR allows to investigate a large number of samples with high precision \cite{ApperleyHH2012}. Spin relaxation is difficult to treat theoretically \cite{LevittDB1992}, which is why new formalisms are required for accurate treatment of this phenomenon in a larger class of situations \cite{KivelsonO1974}. The Mori-Zwanzig formalism is a common and useful tool in this area \cite{Bouchard2007,KivelsonO1974}. However, since modern experiments frequently involve rapidly varying external fields \cite{Bouchard2007,LevittDB1992}, it is very important to have tools available that allow to treat time-dependent Hamiltonians \cite{Bouchard2007}. Due to the success of applying the Mori-Zwanzig theory to spin systems with time-independent Hamiltonians, an extension towards the time-dependent case is a promising path to go.

Spin relaxation in the presence of a time-dependent external magnetic field $\vec{B}(t)$ is described by the famous Bloch equations \cite{KivelsonO1974,Bloch1946,Pule1974}
{\begin{align}%
\tdif{}{t}\bar{S}_z(t) &= \gamma \big(\vec{S}(t) \times \vec{B}(t)\big)_z - \tau_1^{-1}\big(\bar{S}_z(t) - \langle S_z \rangle_{\mathrm{eq}}\big),
\label{zcomponent}\raisetag{5.5ex}\\
\begin{split}%
\tdif{}{t}\bar{S}_\pm (t) &= \gamma \big(\vec{S}(t) \times \vec{B}(t)\big)_x  
\pm \ii \gamma \big(\vec{S}(t) \times \vec{B}(t)\big)_y \\
&\quad\:\!+ \big(\pm\, \ii \omega_0 - \ii \sigma_{\pm} - \tau_2^{-1}\big) \bar{S}_\pm(t),
\end{split}\label{xycomponent}%
\end{align}}%
where
\begin{equation}
\vec{S} = \sum_{i} \vec{S}_i
\end{equation}
is the total spin of the system and the $\{\vec{S}_i\}$ are the individual spins. Overbars denote an instantaneous nonequilibrium average, whereas $\langle \cdot \rangle_{\mathrm{eq}}$ gives the value in thermodynamic equilibrium. The $x$ and $y$ elements of $\vec{S}$ have been re-written in terms of the spin ladder operators
\begin{equation}
S_{\pm} = S_x \pm \ii S_y
\end{equation}
with the property $S_+^\dagger = S_-$. The constants $\tau_1$ and $\tau_2$ are relaxation times that describe how rapidly the system returns to its equilibrium magnetization after an external perturbation, $\gamma$ is the gyromagnetic ratio, $\omega_0$ is the Larmor frequency, and $\sigma_{\pm}$ is the dynamic frequency shift \cite{KivelsonO1974}. In usual NMR terminology, \cref{zcomponent} is referred to as \textit{spin-lattice relaxation} or \textit{longitudinal relaxation}, while \cref{xycomponent} describes \textit{spin-spin relaxation} or \textit{transverse relaxation} \cite{ApperleyHH2012}. The first term on the right-hand side of \cref{zcomponent} and the first two terms on the right-hand side of \cref{xycomponent} are usually not considered in treatments of spin relaxation based on the Mori theory \cite{KivelsonO1974, Bouchard2007}. They are a consequence of the precession of magnetic moments in external magnetic fields, which is, assuming no relaxation, governed by \cite{Bloch1946}
\begin{equation}
\frac{\dif \vec{S}}{\dif t} = \gamma \vec{S} \times \vec{B}.
\end{equation}

The remaining terms on the right-hand side of \cref{zcomponent,xycomponent} describe the relaxation of the total spin towards its equilibrium value. For the purpose of demonstrating the formalism without getting lost in technical difficulties, we set up our example as simple as possible. First, we therefore assume that we are in the regime close to thermal equilibrium in which \cref{bouchard} holds. The magnetic field is assumed to have the form
\begin{equation}
\vec{B}(t) = B(t) \vec{e}_z,
\end{equation}
i.e., it always points in the $z$ direction while having a time-varying amplitude $B(t)$. Our Hamiltonian reads
\begin{equation}
\HS(t) = \omega_0 S_z + \HL + \HSL - \gamma B(t)S_z.
\label{SpinHamiltonian}%
\end{equation}
The first three terms are typical parts of a spin relaxation Hamiltonian, consisting of a spin part (we here use $\omega_0 S_z$, describing Zeeman interaction), a lattice Hamiltonian $\HL$ involving all interactions that commute with the spin Hamiltonian, such as molecular tumbling \cite{KivelsonO1974}, and a spin-lattice Hamiltonian $\HSL$ that describes interactions of spin and lattice and is therefore relevant for the description of the relaxation \cite{KivelsonO1974}. The Hamiltonian can involve a variety of effects such as shielding anisotropy and dipolar coupling to surrounding atoms \cite{ApperleyHH2012}. The last term is the usual Hamiltonian for the case of a spin in a magnetic field pointing in the $z$ direction \cite{Muenster2010}.

For obtaining the Bloch equations, we follow the procedure described, e.g., by Kivelson and Ogan \cite{KivelsonO1974}, but with the important modification that we include a time-dependent external magnetic field and therefore use the Bouchard equation \eqref{bouchard} rather than the usual Mori equation \eqref{mori}. In doing so, we will also make use of the generalized Markovian approximation derived in \cref{GeneralizedMarkovianApproximation}.

We choose the set of slow variables
\begin{equation}
\vec{A} =
\begin{pmatrix}
S_+\\
\Delta S_z\\
S_- \\
\end{pmatrix}
\end{equation}
with 
\begin{equation}
\Delta S_z (t) = S_z (t) - \langle S_z \rangle_{\mathrm{eq}}.
\end{equation}
In Appendix \ref{appendixb}, we explain how the non-Hermitian operators $S_+$ and $S_-$ can be accommodated in the Mori-Zwanzig formalism even though its derivation mostly assumes Hermitian operators.
Furthermore, we make use of the common high-temperature approximation \cite{Grabert1982}. In this limit, the Mori product $(X,Y)_M$ agrees with the thermal average $\langle XY^{\dagger}\rangle_{\mathrm{eq}}$. Using $\langle S_\pm S_\mp\rangle_{\mathrm{eq}}=N\hbar^2/2$ and $\langle S_z S_z\rangle_{\mathrm{eq}}=N\hbar^2/4$, the normalization matrix to lowest order in $\beta$ then reads \cite{KivelsonO1974}
\begin{equation}
\langle \vec{A} \vec{A}^\dagger \rangle_{\mathrm{eq}} = \frac{1}{4} N \hbar^2
\begin{pmatrix}
2 & 0 & 0\\
0 & 1 & 0\\
0 & 0 & 2\\
\end{pmatrix}
\end{equation}
with the number of spins $N$. Next, we require $\Omega_{ij}(t)$, given by \cref{seconddefinition}, for which we need the time derivative
\begin{equation}
\begin{split}%
\ii \LS(t)\vec{A} &= \frac{\ii}{\hbar}[\HS(t), \vec{A}] = \frac{\ii}{\hbar}\omega_0[S_z, \vec{A}] + \frac{\ii}{\hbar}[\HSL, \vec{A}] \\
&\quad\:\! - \gamma B(t)\frac{\ii }{\hbar} [S_z, \vec{A}].
\end{split}\label{commut}%
\end{equation}
The relevant variables commute with $\HL$, so this part of the Hamiltonian does not contribute to \cref{commut}. As the spin ladder operators are not Hermitian, we have to take the Hermitian adjoint of the second operator appearing in a Mori product. For the components $\langle \dot{S}_+ S_- \rangle_{\mathrm{eq}}$, $\langle \dot{S}_- S_+ \rangle_{\mathrm{eq}}$, and $\langle \dot{S}_z S_z \rangle_{\mathrm{eq}}$, we obtain
\begin{align}
\begin{split}
\langle \dot{S}_+ S_- \rangle_{\mathrm{eq}} &= \ii (\omega_0 - \gamma B(t))\langle S_+ S_-\rangle_{\mathrm{eq}} \\
&\quad\:\! + \frac{\ii}{\hbar}\langle [\HSL, S_+] S_- \rangle_{\mathrm{eq}},
\end{split}\\
\begin{split}
\langle \dot{S}_- S_+ \rangle_{\mathrm{eq}} &= - \ii (\omega_0 - \gamma B(t))\langle S_- S_+\rangle_{\mathrm{eq}} \\
&\quad\:\! + \frac{\ii}{\hbar}\langle [\HSL, S_-] S_+ \rangle_{\mathrm{eq}},
\end{split}\\
\begin{split}
\langle \dot{S}_z S_z \rangle_{\mathrm{eq}} &= \frac{\ii}{\hbar} \langle [\HSL, S_z] S_z \rangle_{\mathrm{eq}},
\end{split}
\end{align}
where we have used the standard commutation relation
\begin{equation}
[S_z, S_\pm] = \pm \hbar S_\pm.
\end{equation}
Using in addition the fact that ensemble averages of terms linear in $\HSL$ vanish in most cases that are of practical relevance \cite{KivelsonO1974}, we arrive at
\begin{equation}
\Omega (t) = \ii (\omega_0 - \gamma B(t))
\begin{pmatrix}
1 & 0 & 0 \\
0 & 0 & 0\\
0 & 0 & -1\\
\end{pmatrix}.
\label{ommega}%
\end{equation}
An important difference between transversal and longitudinal relaxation is that the time-dependent part $-\gamma B(t)S_z$ of the Hamiltonian \eqref{SpinHamiltonian} commutes with $S_z$, but not with $S_\pm$. Therefore, the Liouvillian acting on $S_z$, given by
\begin{equation}
\ii \LSO S_z = \frac{\ii}{\hbar}[\HSL, S_z],
\end{equation}
is time-independent, whereas the Liouvillian acting on $S_\pm$, given by
\begin{equation}
\begin{split}
\ii \LS(t) S_\pm &= \frac{\ii}{\hbar}[\HS(t), S_\pm]\\ 
&= \pm\, \ii (\omega_0 - \gamma B(t)) S_\pm + \frac{\ii}{\hbar}[\HSL, S_\pm], 
\end{split}
\end{equation}
is not. Since with $\LSO$ and $\LS(t)$ all relevant Liouvillians commute, we can use ordinary rather than time-ordered exponentials for our calculations.

The random force reads in the Schr\"odinger picture
\begin{equation}
\begin{split}
\vec{F} &= Q \ii \LS(t) \vec{A} = \ii \LS(t) \vec{A} - \Omega (t) \vec{A} \\
&= \frac{\ii}{\hbar} [(\omega_0 - \gamma B(t))S_z + \HSL, \vec{A}] - \Omega (t) \vec{A}\\ 
&= \frac{\ii}{\hbar} [\HSL, \vec{A}].
\end{split}
\end{equation}
Here, we used Eqs.\ \eqref{prolin}, \eqref{seconddefinition}, \eqref{commut}, and \eqref{ommega}. The noise is thus a consequence of spin-lattice interactions that we do not model explicitly.

Now, we have at our hands everything we need to calculate the terms in \cref{bouchard}. We start with
\begin{equation}
\begin{split}
\vec{F}(t,s) &= \exp_R\!\bigg(\ii \INT{s}{t}{t'} Q \LS(t')\bigg)Q\ii \LS(t) \vec{A} \\
&= \frac{\ii}{\hbar} \exp_R\!\bigg(\ii \INT{s}{t}{t'} Q \LS(t')\bigg) [\HSL, \vec{A}].
\end{split}\raisetag{8ex}
\end{equation}
Bouchard's memory kernel \eqref{phiijBouchard} reads here
\begin{equation}
\begin{split}
\mathcal{H}(t,s) &= \langle (\ii \LS(s)\vec{F}(t,s)) \vec{A}^\dagger \rangle_{\mathrm{eq}} \langle \vec{A} \vec{A}^\dagger \rangle_{\mathrm{eq}}^{-1} \\
&= -\langle\vec{F}(t,s)\vec{F}^\dagger \rangle_{\mathrm{eq}} \langle \vec{A} \vec{A}^\dagger \rangle_{\mathrm{eq}}^{-1}.
\end{split}\label{random}%
\end{equation}
To justify \cref{random}, note that \cref{trace} implies
\begin{equation}
\langle (\ii \LS(s) \vec{F}(t,s)) \vec{A}^\dagger \rangle_{\mathrm{eq}} = -\langle \vec{F}(t,s)(\ii \LS(s)\vec{A})^\dagger \rangle_{\mathrm{eq}}.
\end{equation}
We can then expand
\begin{equation}
\begin{split}
&\langle \vec{F}(t,s) (\ii \LS(s)\vec{A})^\dagger \rangle_{\mathrm{eq}} 
= \langle \vec{F}(t,s)((P + Q)\ii \LS(s)\vec{A})^\dagger \rangle_{\mathrm{eq}} \\
&\!= \langle \vec{F}(t,s)(P \ii \LS(s)\vec{A})^\dagger\rangle_{\mathrm{eq}} + \langle \vec{F}(t,s)(Q \ii \LS(s)\vec{A})^\dagger\rangle_{\mathrm{eq}} \\
&\!= \langle \vec{F}(t,s) \vec{F}^{\dagger} \rangle_{\mathrm{eq}}.
\end{split}\raisetag{3ex}%
\end{equation}
Performing the matrix multiplications in \cref{random} gives the elements
{\begin{align}%
\begin{split}%
\mathcal{H}_{11}(t,s) &= -\frac{2}{N}\bigg\langle \frac{1}{\hbar^4} \bigg( \! \exp_R\!\bigg(\ii \INT{s}{t}{t'}Q \LS(t')\bigg) [\HSL, S_+] \bigg) \\ 
&\qquad\quad\; [\HSL, S_+]^\dagger \bigg\rangle_{\mathrm{eq}},  
\end{split}\raisetag{7ex}\\
\begin{split}%
\mathcal{H}_{22}(t,s) &= -\frac{4}{N}\bigg\langle \frac{1}{\hbar^4} \bigg( \! \exp_R\!\bigg(\ii \INT{s}{t}{t'}Q \LS(t')\bigg) [\HSL, S_z] \bigg) \\ 
&\qquad\quad\; [\HSL, S_z]^\dagger \bigg\rangle_{\mathrm{eq}},  
\end{split}\raisetag{7ex}\\
\begin{split}%
\mathcal{H}_{33}(t,s) &= -\frac{2}{N}\bigg\langle \frac{1}{\hbar^4} \bigg( \! \exp_R\!\bigg(\ii \INT{s}{t}{t'}Q \LS(t')\bigg) [\HSL, S_-] \bigg) \\ 
&\qquad\quad\; [\HSL, S_-]^\dagger \bigg\rangle_{\mathrm{eq}}. 
\end{split}\raisetag{7ex}%
\end{align}}%
Due to the typical symmetries of $\HSL$, we have $\mathcal{H}_{ij}(t,s)=0$ for $i\neq j$ \cite{KivelsonO1974}. This leads to a decoupling of the equations for the relevant variables. Using $[\HSL, S_z]^\dagger = - [\HSL,S_z]$ and defining $F_z(t,0)=F_2(t,0)$, we arrive at
\begin{equation}
\begin{split}%
\tdif{}{t}S_z(t) 
&= \frac{4}{N} \INT{0}{t}{s} \bigg\langle \frac{1}{\hbar^4} \bigg( \! \exp_R\!\bigg(\ii \INT{s}{t}{t'}Q \LS(t')\bigg) [\HSL, S_z] \bigg)  \\
&\qquad\qquad\quad\;\, [\HSL, S_z] \bigg\rangle_{\mathrm{eq}}(S_z (s) - \langle S_z\rangle_{\mathrm{eq}})  \\
&\quad\, + F_z(t,0)\\
&= \frac{4}{N} \INT{0}{t}{s} \bigg\langle \frac{1}{\hbar^4} \bigg( \! \exp_R\!\bigg(\ii \INT{t-s}{t}{t'}Q \LS(t')\bigg) [\HSL, S_z] \bigg)  \!\!\!\! \\
&\qquad\qquad\quad\;\, [\HSL, S_z] \bigg\rangle_{\mathrm{eq}}  (S_z (t-s) - \langle S_z\rangle_{\mathrm{eq}}) \\
&\quad\, + F_z(t,0),
\end{split}\label{spinz}\raisetag{20ex}%
\end{equation}
where we have made the substitution $s \to t-s$ and switched the integration boundaries.

We can now take the instantaneous nonequilibrium average, so that the random force $F_z(t,0)$ vanishes, and make a Markovian approximation, which here just implies that $\bar{S}_z(t-s)$ can be replaced by $\bar{S}_z(t)$ with the upper integration limit extended to infinity. 
Moreover, as $\ii P \LS(t)X$ is of order $\ii \LS(t)A_i$, we can replace $\exp_R(\ii \TINT{t-s}{t}{t'} Q \LS(t'))$ by $\exp_R(\ii \TINT{t-s}{t}{t'} \LS(t'))$ in this approximation. 
This holds, since $\ii P \LS(t)X = A_j (A_j, A_k)_M^{-1} (\ii \LS(t)X, A_k)_M = -A_j (A_j, A_k)_M^{-1} (X, \ii \LS(t) A_k)_M$ \cite{Zwanzig2001}. 

Therefore, the result of \cref{spinz} can be written as
\begin{equation}
\tdif{}{t} \bar{S}_z(t) = -\tau_1^{-1}(t) \big(\bar{S}_z(t) - \langle S_z \rangle_{\mathrm{eq}}\big),
\label{dsz}%
\end{equation}
with the inverse relaxation time
\begin{equation}
\begin{split}
\tau_1^{-1}(t) &= \frac{4}{N} \INT{0}{\infty}{s} \bigg\langle \frac{1}{\hbar^4} \bigg(\!\exp_R\!\bigg(\ii \INT{t-s}{t}{t'} \LS(t')\bigg)[\HSL, S_z]\bigg) \\
&\qquad\qquad\quad\;\, [S_z, \HSL] \bigg\rangle_{\mathrm{eq}}.
\end{split}\label{tau1}\raisetag{7ex}%
\end{equation}
The structure of Eqs.\ \eqref{dsz} and \eqref{tau1} for the longitudinal relaxation does not differ from the standard result \cite{KivelsonO1974}. However, the relaxation time \eqref{tau1} now has explicit time dependence. As discussed in \cref{GeneralizedMarkovianApproximation}, this is a result of applying the Markovian approximation in the case of time-dependent Hamiltonians. The time dependence might be surprising, since the magnetic field points in the $z$ direction and should therefore not influence the motion of $S_z$. This issue is resolved by the fact that the spin-lattice Hamiltonian $\HSL$ is typically constructed using spin operators \cite{KivelsonO1974}, which can lead to a coupling of $S_z$ to $S_\pm$ and thus to the magnetic field. Whether this will occur and how strong the effect is depends on the exact form of $\HSL$. 

Now, we come to the more interesting part. The expression for $S_\pm(t)$ reads
\begin{equation}
\begin{split}
S_\pm (t) &= \exp\!\bigg(\ii \INT{0}{t}{t'} \LS(t')\bigg) S_\pm \\
&= \exp\!\bigg(\!\pm \ii \omega_0 t \mp \ii \gamma \INT{0}{t}{t'} B(t') + \ii \LSO t\bigg) S_\pm.
\end{split}\raisetag{8ex}%
\end{equation}
To simplify the expression, we write $b(t) = \TINT{0}{t}{t'} B(t')$, giving
\begin{equation}
\begin{split}
S_\pm (t) = \exp(\pm\, \ii \omega_0 t \mp \ii \gamma b(t) + \ii \LSO t) S_\pm. 
\end{split}
\end{equation}
We use the identity $[\HSL, S_\pm]^\dagger = - [\HSL,S_\mp]$. The equation of motion for $S_{\pm}(t)$ then reads
\begin{equation}
\begin{split}
\tdif{}{t} S_\pm (t)
&= \pm\, \ii (\omega_0 - \gamma B(t)) S_\pm(t)\\ 
&\!\!\!\!\!\!\!\!\!\!\!\! \quad\:\! + \frac{2}{N} \INT{0}{t}{s} 
\bigg\langle \frac{1}{\hbar^4} \bigg( \! \exp_R\!\bigg(\ii \INT{s}{t}{t'}Q \LS(t')\bigg) [\HSL, S_\pm] \bigg) \\ 
&\!\!\!\!\!\!\!\!\!\!\!\! \qquad\qquad\qquad\;\:\! [\HSL, S_\mp] \bigg\rangle_{\mathrm{eq}} \\
&\!\!\!\!\!\!\!\!\!\!\!\! \qquad\qquad\quad\;\,\:\! \exp(\pm\, \ii \omega_0 s \mp \ii \gamma b(s) + \ii \LSO s) S_\pm \\
&\!\!\!\!\!\!\!\!\!\!\!\! \quad\:\! + F_\pm(t,0)
\end{split}\label{spinac}\raisetag{13ex}%
\end{equation}
with $F_+(t,0)=F_1(t,0)$ and $F_-(t,0)=F_3(t,0)$. 
A remarkable aspect of this result is the term $\mp\, \ii \gamma B(t) S_\pm (t)$. To see its significance, notice that
\begin{equation}
\begin{pmatrix}
S_x\\
S_y\\
S_z
\end{pmatrix}
= \frac{1}{2}
\begin{pmatrix}
S_+ + S_-\\
-\ii (S_+ - S_-)\\
2 S_z
\end{pmatrix}
\end{equation}
and that, when dropping the memory and noise terms in \cref{spinz,spinac} and setting $\omega_0 = 0$ for a moment, the term $\mp\, \ii \gamma B(t) S_\pm(t)$ leads to
\begin{equation}
\begin{split}
\tdif{}{t}\!
\begin{pmatrix}
S_x\\
S_y\\
S_z
\end{pmatrix}
&= \frac{1}{2}
\begin{pmatrix}
\dot{S}_+ + \dot{S}_-\\
-\ii (\dot{S}_+ - \dot{S}_-)\\
2 \dot{S}_z
\end{pmatrix}\\
&= - \frac{\ii}{2} \gamma B(t)
\begin{pmatrix}
S_+ - S_- \\
-\ii(S_+ + S_-)\\
0
\end{pmatrix}\\
&= - \frac{\ii}{2} \gamma B(t)
\begin{pmatrix}
S_x + \ii S_y - S_x + \ii S_y \\
-\ii(S_x + \ii S_y + S_x - \ii S_y)\\
0
\end{pmatrix}\\
&
= \gamma B(t)
\begin{pmatrix}
S_y \\
- S_x\\
0
\end{pmatrix}
= \gamma B(t) \vec{S} \times \vec{e}_z \\
&= \gamma \vec{S} \times \vec{B}(t).
\end{split}\raisetag{25ex}%
\end{equation}
The latter expression corresponds to the coupling term that is given in the complete Bloch equations \eqref{zcomponent} and \eqref{xycomponent}. Therefore, we can model the coupling to time-dependent external magnetic fields. Note, however, that an analogous result could also have been obtained in the standard formalism for the case of a time-independent magnetic field $\vec{B}_0$, which would have resulted in $\vec{S} \times \vec{B}_0$ instead of $\vec{S} \times \vec{B}(t)$.

Equation \eqref{spinac} can be approximated using a procedure in analogy to a transformation described, e.g., by Kivelson and Ogan for the simpler case of no time-dependent fields \cite{KivelsonO1974}. It goes slightly beyond the usual Markovian approximation presented in \cref{GeneralizedMarkovianApproximation}. Although our variables decay slowly, they might be subject to fast oscillations due to Larmor precession and a rapidly varying external magnetic field. Referring to \cref{bouchard}, this fast part of the dynamics is represented in the frequency matrix $\Omega(t)$, i.e.,
\begin{equation}
\frac{\dif}{\dif t} \vec{A}(t) = \Omega(t) \vec{A}(t) + \text{slow terms}.
\end{equation}
This equation is formally solved by
\begin{equation}
\vec{A}(t) = \exp\!\bigg(\INT{0}{t}{t'} \Omega(t')\bigg) \vec{A}.
\end{equation}
Therefore, we can assume
\begin{equation}
\exp\!\bigg(\!-\INT{0}{t}{t'} \Omega(t')\bigg) \vec{A}(t)
\end{equation}
to be a slowly varying quantity that changes only due to the slow relaxation. Since $\Omega_{22}=0$, the dynamics of $S_z$ is not affected by this transformation, which justifies our previous treatment.
A time-ordered exponential is not necessary here, since $\Omega(t)$ is a diagonal matrix containing numbers rather than operators that will not cause any problems related to noncommutativity.
We now write
\begin{equation}
\INT{0}{t}{s} \mathcal{H}(t,s) \bar{\vec{A}}(s) = \INT{0}{t}{s} \mathcal{H}_{\mathrm{eff}}(t,s) \exp\!\bigg(\!-\INT{0}{s}{t'} \Omega(t')\bigg) \bar{\vec{A}}(s),
\label{eff}%
\end{equation}
defining the effective memory function
\begin{equation}
\mathcal{H}_{\mathrm{eff}}(t,s) = \mathcal{H}(t,s) \exp\!\bigg(\INT{0}{s}{t'} \Omega(t')\bigg)
\end{equation}
that will decay very rapidly. This allows to replace $\exp(-\TINT{0}{s}{t'} \Omega(t')) \bar{\vec{A}}(s)$ by $\exp(-\TINT{0}{t}{t'} \Omega(t')) \bar{\vec{A}}(t)$ and to increase the upper integration limit of the second integral in \cref{eff} to infinity. The resulting dynamic equation for $\bar{\vec{A}}(t)$ is
\begin{equation}
\tdif{}{t} \bar{\vec{A}}(t) = (\Omega (t) + \mathcal{H}_R(t) + \ii \mathcal{H}_I(t)) \bar{\vec{A}}(t),
\end{equation}
where 
\begin{equation}
\mathcal{H}_R(t) + \ii \mathcal{H}_I(t) = \INT{0}{\infty}{s} \mathcal{H}_{\mathrm{eff}}(t,s) \exp\!\bigg(\!-\INT{0}{t}{t'} \Omega(t')\bigg)
\end{equation}
is the relaxation matrix decomposed into real and imaginary parts. 

As a consequence of the aforementioned considerations, we assume
\begin{equation}
\exp\!\bigg(\!\mp\, \ii \omega_0t \pm \INT{0}{t}{t'} \ii \gamma B(t')\bigg) S_\pm(t)
\end{equation}
to be a slowly varying quantity. 
Inserting 
\begin{equation}
\begin{split}%
1 = \,&\exp\!\bigg(\!\pm\, \ii \omega_0s \mp \INT{0}{s}{t'} \ii \gamma B(t')\bigg) \\
&\exp\!\bigg(\!\mp\, \ii \omega_0s \,\pm \INT{0}{s}{t'} \ii \gamma B(t')\bigg)
\end{split}\label{one}%
\end{equation}
into \cref{spinac} gives
\begin{equation}
\begin{split}
\tdif{}{t} S_\pm(t) &= \pm \ii (\omega_0 - \gamma B(t)) S_\pm(t) \\
&\!\!\!\!\!\!\!\!\!\!\!\!\quad\:\! + \frac{2}{N} \INT{0}{t}{s} \bigg\langle \frac{1}{\hbar^4} \bigg(\! \exp_R\!\bigg(\ii \INT{s}{t}{t'}Q \LS(t')\bigg) [\HSL, S_\pm] \bigg) \\
&\!\!\!\!\!\!\!\!\!\!\!\! \qquad\qquad\qquad\;\:\! [\HSL, S_\mp]\bigg\rangle_{\mathrm{eq}}\\
&\!\!\!\!\! \qquad\qquad\;\:\! \exp\!\bigg(\!\pm\, \ii \omega_0s \,\mp \INT{0}{s}{t'} \ii \gamma B(t')\bigg) \\
&\!\!\!\!\! \qquad\qquad\;\:\! \exp\!\bigg(\!\mp\, \ii \omega_0s \,\pm \INT{0}{s}{t'} \ii \gamma B(t')\bigg) S_\pm(s) \\
&\!\!\!\!\!\!\!\!\!\!\!\!\quad\:\! + F_\pm (t,0).
\end{split}\raisetag{18ex}%
\end{equation}
Again, we use the shorthand notation $b(t) = \TINT{0}{t}{t'} B(t')$. The quasi-stationary approximation corresponds to replacing  the slow quantity by its value at $s=t$ and $\exp_R(\ii \INT{t-s}{t}{t'}Q \LS(t'))$ by $\exp_R(\ii \INT{t-s}{t}{t'} \LS(t'))$. Applying additionally an instantaneous nonequilibrium average to get an expression for mean values rather than operators and to remove the random force term yields
\begin{equation}
\begin{split}
\tdif{}{t} \bar{S}_\pm (t) &= \pm \ii (\omega_0 - \gamma B(t)) \bar{S}_\pm(t) \\
&\!\!\!\!\!\!\!\!\!\!\!\!\quad\:\! + \frac{2}{N} \INT{0}{t}{s} \bigg\langle \frac{1}{\hbar^4} \bigg(\! \exp_R\!\bigg(\ii \INT{s}{t}{t'} \LS(t')\bigg) [\HSL, S_\pm]\bigg) \\
&\!\!\!\!\!\!\!\!\!\!\!\! \qquad\qquad\qquad\;\:\! [\HSL, S_\mp]\bigg\rangle_{\mathrm{eq}}\\ 
&\!\!\!\!\! \qquad\qquad\;\:\! \exp(\pm\, \ii \omega_0s \mp \ii \gamma b(s)) \\ 
&\!\!\!\!\! \qquad\qquad\;\:\! \exp(\mp\,\ii \omega_0 t \pm \ii \gamma b(t)) \bar{S}_\pm(t).
\end{split}\raisetag{13ex}%
\end{equation}
In our general discussion in \cref{GeneralizedMarkovianApproximation}, we have explained how to derive the approximated kernel by extending the integration in $s$ to infinity and making a series of substitutions. 
Those can now be applied to the problem at hand, i.e., to the term
\begin{equation}
\begin{split}
&\frac{2}{N} \INT{0}{t}{s} \bigg\langle \frac{1}{\hbar^4} \bigg(\! \exp_R\!\bigg(\ii \INT{s}{t}{t'} \LS(t')\bigg) [\HSL, S_\pm] \bigg) [\HSL, S_\mp]\bigg\rangle_{\mathrm{eq}}\\
&\qquad\quad\, \exp(\pm\, \ii \omega_0s \mp \ii \gamma b(s)) \exp(\mp \,\ii\omega_0t \pm \ii \gamma b(t)) .
\end{split}\raisetag{3ex}%
\end{equation}
We extend the lower integration limit to $-\infty$, as the kernel is negligible for large $t-s$ anyway, and switch the integration boundaries. Substituting afterwards $s$ by $t-s$ leads to 
\begin{equation}
\begin{split}
&\approx \frac{2}{N} \INT{0}{\infty}{s} \bigg\langle \frac{1}{\hbar^4} \bigg( \! \exp_R\!\bigg(\ii \INT{t-s}{t}{t'} \LS(t')\bigg) [\HSL, S_\pm] \bigg) [\HSL, S_\mp]\bigg\rangle_{\mathrm{eq}}\\
&\qquad\qquad\;\; \exp(\mp\, \ii \omega_0s \pm \ii \omega_0 t \mp \ii \gamma b(t-s)) \\ 
&\qquad\qquad\;\; \exp(\mp\, \ii \omega_0 t \pm \ii \gamma b(t)).
\end{split}\raisetag{6ex}%
\end{equation}
We let the factors $e^{\pm\,\ii \omega_0 t}$ cancel, which yields
\begin{equation}
\begin{split}
&= \frac{2}{N} \INT{0}{\infty}{s} \bigg\langle \frac{1}{\hbar^4} \bigg( \! \exp_R\!\bigg(\ii \INT{t-s}{t}{t'} \LS(t')\bigg) [\HSL, S_\pm] \bigg) [\HSL, S_\mp]\bigg\rangle_{\mathrm{eq}}\\
&\qquad\qquad\;\; \exp(\mp\,\ii \omega_0s \mp \ii \gamma b(t-s)) \exp(\pm \ii \gamma b(t)).
\end{split}\raisetag{3ex}%
\end{equation}
Making the definitions
\begin{equation}
\begin{split}
\tau_{2 \pm}^{-1}(t) &= \Rea\!\bigg( \frac{2}{N} \INT{0}{\infty}{s} \bigg\langle \frac{1}{\hbar^4} \bigg( \! \exp_R\!\bigg(\ii \INT{t-s}{t}{t'} \LS(t')\bigg) [\HSL, S_\pm] \bigg) \\
&\qquad\quad\; [S_\mp, \HSL]\bigg\rangle_{\mathrm{eq}} \!\!\!\!\! \exp(\mp\, \ii \omega_0s \mp \ii \gamma b(t-s) \pm \ii \gamma b(t))\!\bigg)
\end{split}\label{tau2}%
\end{equation}
and 
\begin{equation}
\begin{split}
\sigma_\pm (t) &= \Ima\!\bigg( \frac{2}{N} \INT{0}{\infty}{s} \bigg\langle \frac{1}{\hbar^4} \bigg( \! \exp_R\!\bigg(\ii \INT{t-s}{t}{t'} \LS(t')\bigg) [\HSL, S_\pm] \bigg) \\
&\qquad\quad\;\:\! [S_\mp, \HSL]\bigg\rangle_{\mathrm{eq}} \!\!\!\!\! \exp(\mp\, \ii \omega_0s \mp \ii \gamma b(t-s) \pm \ii \gamma b(t))\!\bigg),
\end{split}\label{sigmapm}%
\end{equation}
where $\Rea$ and $\Ima$ denote real and imaginary parts, respectively, we can finally write the equation of motion for $\bar{S}_{\pm}(t)$ as
\begin{equation}
\begin{split}
\tdif{}{t} \bar{S}_\pm (t) &= \big(\!\pm\, \ii \omega_0 \mp \ii \gamma B(t) - \big(\ii\sigma_\pm (t) + \tau_{2 \pm}^{-1}(t)\big)\big) \bar{S}_\pm (t).
\end{split}\label{finaleqofmotion}%
\end{equation}

A few points are notable about this result. First, as explained above, we have obtained the coupling to the magnetic field in the form that is well known from experiments and other derivations. Second, with $\sigma_{\pm}$ and $\tau_{2 \pm}$ we again have explicitly time-dependent transport coefficients. This effect will be stronger than for $\tau_1$ (see \cref{tau1}), since here we have a direct coupling to the magnetic field, which leads to the factor $\exp(\mp\, \ii \omega_0s \mp \ii \gamma b(t-s) \pm \ii \gamma b(t))$ in \cref{tau2,sigmapm}. 

For a time-independent magnetic field $B_0$, the exponentials in \cref{tau2,sigmapm} would give
\begin{equation}
\begin{split}
&\:\! \exp(\mp\, \ii \omega_0s \mp \ii \gamma b(t-s) \pm \ii \gamma b(t))\\
&=\exp(\mp\, \ii \omega_0s \mp \ii \gamma B_0 (t-s) \pm \ii \gamma B_0 t)\\
&= \exp(\mp\, \ii \omega_0s \pm \ii \gamma B_0 s).
\end{split}
\end{equation}
Since $s$ is integrated over and $\exp_R(\ii \TINT{t-s}{t}{t'} \LS(t'))$ could be replaced by $\exp(\ii \LS s)$, this would lead to time-independent coefficients. Equation \eqref{tau2} is then basically identical with Kivelson's and Ogan's result for the relaxation time \cite{KivelsonO1974}, the only difference being that $\omega_0$ has been replaced by $\omega_0 - \gamma B_0$ in our case. 

We can also obtain equations for the time evolution of the correlation functions, which here correspond to averages of products of the relevant quantities at different times. The general evolution law is given by \cref{simplecorrelation}. In the scenario at hand, the relevant matrices $\Omega_{ij}$ and $\mathcal{H}_{ij}$ are diagonal. This is why we get three decoupled differential equations for the diagonal entries of the correlation function matrix, whereas other entries are constant. The relevant elements of the correlation function matrix
\begin{equation}
C (t,0) = \langle \vec{A}(t) \vec{A}^\dagger(0) \rangle_{\mathrm{eq}}
\end{equation}
are
{\begin{align}%
C_{11}(t,0) &= \langle S_+(t) S_- \rangle_{\mathrm{eq}},\\
C_{22}(t,0) &= \langle \Delta S_z(t) \Delta S_z \rangle_{\mathrm{eq}},\\
C_{33}(t,0) &= \langle S_-(t) S_+ \rangle_{\mathrm{eq}}.
\end{align}}%
As discussed in \cref{approximations}, all approximations used for the observables themselves are also valid for the correlation functions. We can thus immediately write
{\begin{align}%
\begin{split}%
\tdif{}{t} C_{11}(t,0) &= \Big(\ii \omega_0 - \ii \gamma B(t) \\
&\quad\;\;\, -\big(\ii \sigma_+ (t) + \tau_{2+}^{-1}(t)\big)\Big)  C_{11}(t,0),
\end{split}\raisetag{7ex}\\
\tdif{}{t} C_{22}(t,0) &= - \tau_1^{-1}(t) C_{22}(t,0),\\
\begin{split}%
\tdif{}{t} C_{33}(t,0) &= \Big(\! - \ii \omega_0 + \ii \gamma B(t) \\
&\quad\;\;\,\:\! -\big(\ii \sigma_- (t) + \tau_{2-}^{-1}(t)\big) \Big) C_{33}(t,0),
\end{split}\raisetag{7ex}%
\end{align}}%
where all constants are defined as above.

\section{\label{conclusions}Conclusions}
We have derived a generalization of the Mori-Zwanzig projection operator formalism that can be applied to systems without and with time-dependent Hamiltonians. It is applicable to classical and quantum-mechanical systems, close to and far from thermodynamic equilibrium, and even for observables with explicit time dependence. Moreover, we have described a variety of approximation methods that allow to considerably simplify the resulting equations in the case that the relevant variables are varying slowly or that the system is close to equilibrium. In the latter situation, known results from the literature arise as a limiting case, which confirms our theory.

Our results are of great importance for a variety of applications, including the derivation of mesoscopic and macroscopic field theories for systems consisting of many particles. Such systems are frequently subject to time-dependent external influences such as periodic driving. Using the extended formalism, it is possible to treat them within the projection operator framework. The use of approximations for slowly relaxing and close-to-equilibrium systems, which are applicable to a large number of situations typically considered, greatly increase the usefulness of the resulting equations.

Possible continuations of this work include the application to specific systems. For example, the discussion of spin relaxation described above can be extended towards stronger magnetic fields, which leads to larger deviations from equilibrium, or more complex spin Hamiltonians. Many other applications are possible as well. Among them are applications to classical systems such as driven soft matter. 
A particularly challenging task for future research would be a further extension of the Mori-Zwanzig formalism towards arbitrary non-Hermitian operators and Hamiltonians.  
Moreover, it is possible to develop additional approximation techniques based, e.g., on Magnus expansions or convolutionless equations, as well as methods for analyzing large-scale fluctuations at critical points and for mappings to non-Hamiltonian dynamical systems.

\begin{acknowledgments}
We thank Jens Bickmann and Louis Bouchard for helpful discussions. 
R.W.\ is funded by the Deutsche Forschungsgemeinschaft (DFG, German Research Foundation) -- WI 4170/3-1. 
\end{acknowledgments}

\appendix
\section{\label{appendixa}Expression for the frequency matrix}
In this appendix, we show by an explicit calculation that choosing the form \eqref{correlation} for the generalized correlator allows to derive the expression \eqref{relation} for the frequency matrix.

We assume $0 \leq s \leq t$. Equation \eqref{correlation} yields 
\begin{equation}
\begin{split}
&(X(t), \dot{A}_j(s))\lambda_j(s)\\
&\!=\INT{0}{1}{\alpha} \Tr\!\bigg(\bar{\rho}(s)\\ 
&\qquad\quad\;\;\:\! \bigg(\exp_L\!\bigg(\!-\ii \INT{0}{s}{t'} \LS(t')\bigg) X(t)\bigg) e^{-\alpha \lambda_k(s) A_k}\\ 
&\qquad\quad\;\;\:\! \bigg(\exp_L\!\bigg(\!-\ii \INT{0}{s}{t'} \LS(t')\bigg) \dot{A}_j(s)\bigg) e^{\alpha \lambda_l(s)A_l}\bigg)\lambda_j(s).
\end{split}\raisetag{13ex}%
\end{equation}
Using \cref{lidentity} and
\begin{equation}
\begin{split}
&\:\!\exp_L\!\bigg(\!-\ii \INT{0}{s}{t'} \LS(t')\bigg) \dot{A}_i(s)\\
&= \exp_L\!\bigg(\!-\ii \INT{0}{s}{t'} \LS(t')\bigg) \exp_R\!\bigg(\ii \INT{0}{s}{t'} \LS(t')\bigg) \ii \LS(s) A_i\\
&= \ii \LS(s) A_i,
\end{split}\raisetag{11ex}%
\end{equation}
we obtain
\begin{equation}
\begin{split}
&(X(t), \dot{A}_j(s))\lambda_j(s) \\
&\!= - \Tr\!\bigg(\!\exp_L\!\bigg(\!-\ii \INT{0}{s}{t'} \LS(t')\bigg) X(t)\ii \LS(s)\bar{\rho}(s)\bigg).
\end{split}\raisetag{7ex}%
\end{equation}
Equation \eqref{trace} allows to move the Liouvillian from the probability density to the variable. This finally gives
\begin{equation}
\begin{split}
&(X(t), \dot{A}_j(s))\lambda_j(s) \\
&\!= \Tr\!\bigg(\bar{\rho}(s) \ii \LS(s) \exp_L\!\bigg(\!-\ii \INT{0}{s}{t'} \LS(t')\bigg) X(t)\bigg).
\end{split}\label{iden}%
\end{equation}
Moreover, using \cite{Grabert1982,Grabert1978}
\begin{equation}
\frac{\partial \bar{\rho}(t)}{\partial \lambda_i(t)} = - \INT{0}{1}{\alpha} e^{-\alpha\lambda_j(t)A_j} \delta A_i e^{\alpha\lambda_k(t)A_k}\bar{\rho}(t)
\label{iden3}%
\end{equation}
we get
\begin{equation}
\begin{split}
&\:\! \Tr\!\bigg(\frac{\partial \bar{\rho}(s)}{\partial \lambda_i(s)} \exp_L\!\bigg(\!-\ii \INT{0}{s}{t'} \LS(t')\bigg) X(t)\bigg) \\
&= \Tr\!\bigg(\!-\INT{0}{1}{\alpha} e^{-\alpha\lambda_j(s)A_j} \delta A_i e^{\alpha\lambda_k(s)A_k} \bar{\rho}(s) \\
&\qquad\qquad\qquad\;\, \exp_L\!\bigg(\!-\ii \INT{0}{s}{t'} \LS(t')\bigg) X(t)\bigg)\\
&= -\INT{0}{1}{\alpha} \Tr\!\bigg(\bar{\rho}(s) \bigg(\exp_L\!\bigg(\!-\ii \INT{0}{s}{t'} \LS(t')\bigg) X(t)\bigg) \\
&\qquad\qquad\qquad\;\, e^{-\alpha\lambda_j(s)A_j} \delta A_i e^{\alpha\lambda_k(s)A_k}\bigg)\\
&= -\INT{0}{1}{\alpha} \Tr\!\bigg(\bar{\rho}(s) \bigg(\exp_L\!\bigg(\!-\ii \INT{0}{s}{t'} \LS(t')\bigg) X(t)\bigg) e^{-\alpha\lambda_j(s)A_j} \\ 
&\qquad\qquad\qquad\; \bigg(\exp_L\!\bigg(\!-\ii \INT{0}{s}{t'} \LS(t')\bigg) \delta A_i(s)\bigg) e^{\alpha\lambda_k(s)A_k}\bigg)\\
&= -(X(t), \delta A_i(s)),
\end{split}\label{iden2}\raisetag{24ex}
\end{equation}
where we have used the invariance of the trace $\Tr$ under cyclic permutations as well as the relation 
\begin{equation}
\delta A_i = \exp_L\!\bigg(\!-\ii \INT{0}{s}{t'} \LS(t')\bigg) \delta A_i(s).
\end{equation}

The identities just proven allow to derive an equation for the frequency matrix $\Omega_{ij}(t)$ in terms of the correlator \eqref{correlation}. We can write the organized drift \eqref{v} as
{\begin{equation}%
\begin{split}%
v_i(t) &= \Tr(\bar{\rho}(t) \ii \LS(t) A_i) \\
&= \Tr\!\bigg(\bar{\rho}(t)\, \ii \LS(t) \exp_L\!\bigg(\!-\ii \INT{0}{t}{t'} \LS(t')\bigg)\\
&\qquad\quad \exp_R\!\bigg(\ii \INT{0}{t}{t'} \LS(t')\bigg) A_i\bigg)\\
&= \Tr\!\bigg(\bar{\rho}(t)  \ii \LS(t) \exp_L\!\bigg(\!-\ii \INT{0}{t}{t'} \LS(t')\bigg) A_i(t)\bigg).
\end{split}\raisetag{13ex}%
\end{equation}}%
Using \cref{iden} leads to
\begin{equation}
v_i(t) = (A_i(t), \dot{A}_j(t)) \lambda_j(t).
\label{vi}%
\end{equation}
Equations \eqref{iden}, \eqref{iden2}, and \eqref{vi} give
\begin{equation}
\frac{\partial v_i(t)}{\partial \lambda_j(t)} = - (\dot{A}_i(t), \delta A_j (t))
\label{partialvl}%
\end{equation}
and
\begin{equation}
\begin{split}
\frac{\partial a_i(t)}{\partial \lambda_j(t)}&= -(A_i(t), \delta A_j(t))\\
&= -(a_i(t) + \delta A_i(t),\delta A_j(t)) \\ 
&= -(\delta A_i(t), \delta A_j(t)).
\end{split}\label{partialal}%
\end{equation}
Combining \cref{Omegaij} with \cref{partialvl,partialal} yields
\begin{equation}
\begin{split}
\Omega_{ij}(t) &= \frac{\partial v_i(t)}{\partial a_j(t)} = \frac{\partial v_i(t)}{\partial \lambda_k(t)} \frac{\partial \lambda_k(t)}{\partial a_j(t)}\\
&= (\delta A_j(t), \delta A_k(t))^{-1}(\dot{A}_i(t), \delta A_k(t)),
\end{split}
\end{equation}
which is the desired result. Note that the derivation becomes slightly more complicated in the case of non-Hermitian operators, since the proof of \cref{iden3} (see Refs. \cite{Grabert1982,Grabert1978}) is based on choosing the form \eqref{rhobar} for $\bar{\rho}(t)$, which has to be modified for non-Hermitian operators.

\section{\label{appendixb}Non-Hermitian spin ladder operators}
Although it is very common to use the non-Hermitian spin ladder operators $S_+$ and $S_-$ in treatments of spin relaxation based on the Mori-Zwanzig formalism \cite{Grabert1982,Bouchard2007,KivelsonO1974}, it is not trivial to justify why this is possible, since derivations of the formalism typically assume Hermitian operators \cite{Grabert1982}. This issue is usually not discussed. Therefore, we explain here how exactly it can be solved.

As mentioned in \cref{poc}, the main problem is the construction of the relevant probability density. If we take the derivative of \cref{macroequivalence} with respect to $a_j(t)$, we obtain \cite{Grabert1982}
\begin{equation}
\Tr\!\bigg(\pdif{\bar{\rho}(t)}{a_j(t)} A_i\bigg)\! 
= \pdif{\lambda_k(t)}{a_j(t)} \Tr\!\bigg(\pdif{\bar{\rho}(t)}{\lambda_k(t)} A_i\bigg)\! 
= \delta_{ij}.
\end{equation}
Consider now the case that our relevant variables are $S_+$ and $S_-$. As in \cref{spins}, we assume that we are close to equilibrium, which here, if we choose the form \eqref{rhobar} for the relevant density, implies that $\bar{\rho}(t) \approx \frac{1}{Z(t)} (1 - \lambda_j(t)A_j)$ with $j\in\{+,-\}$ and $A_\pm=S_\pm$. This leads to\footnote{To verify the second equality in \cref{traces+}, evaluate the traces in this equation in a basis of $S_z$ eigenstates using the usual properties of the spin ladder operators.}
\begin{equation}
\begin{split}
&\!\!\!\!\!\pdif{\lambda_k(t)}{a_+(t)}\Tr\!\bigg(\pdif{\bar{\rho}(t)}{\lambda_k(t)} S_-\bigg)\\
= & - \frac{1}{Z(t)}\pdif{\lambda_+(t)}{a_+(t)}\Tr(S_+ S_-) - \frac{1}{Z(t)}\pdif{\lambda_-(t)}{a_+(t)}\Tr(S_- S_-)\\
= & - \frac{N}{2}\hbar^2 \pdif{\lambda_+(t)}{a_+(t)} = 0
\end{split}\label{traces+}\raisetag{7ex}%
\end{equation}
with $a_\pm(t) = \Tr(\bar{\rho}(t) S_\pm)$. Similar arguments apply, if we use the canonical form \eqref{gc}. Equation \eqref{traces+} shows that, for the usual choice \eqref{rhobar}, the conjugate variable $\lambda_+(t)$ has no dependence on $S_+$. This is not a reasonable definition of a \ZT{conjugate variable}, and it will make calculations more difficult, since the resulting equations for the relevant variables are always coupled. We can avoid this problem, if we write
\begin{equation}
\bar{\rho}(t) = \frac{1}{Z(t)} e^{-\lambda_j(t) A_j^\dagger} 
= \frac{1}{Z(t)} e^{-\lambda_+(t) S_- - \lambda_-(t) S_+}.
\label{rhobarmod}%
\end{equation}
This form reduces to \cref{rhobar} for self-adjoint operators and is therefore a reasonable generalization for the case of spin ladder operators. 
To ensure that $\bar{\rho}(t)$ is Hermitian, $\lambda_+(t)$ and $\lambda_-(t)$ need to be complex conjugates of each other. This has to be taken into account when constructing the entropy or free energy that determines $\lambda_+(t)$ and $\lambda_-(t)$.

The form \eqref{rhobarmod} has the additional advantage that \cref{lidentity} changes to
\begin{equation}
- \ii \LS(t) \bar{\rho}(t) = \lambda_j(t) \INT{0}{1}{\alpha} e^{-\alpha \lambda_k(t) A^\dagger_k} \ii \LS(t) A_j^\dagger e^{\alpha \lambda_l(t) A^\dagger_l}\bar{\rho}(t).
\end{equation}
This identity has been used to derive \cref{relation} for the frequency matrix $\Omega_{ij}(t)$ and \cref{phiijBouchard} for the memory matrix $\mathcal{H}_{ij}(t,s)$. The modification of \cref{lidentity} requires to take the Hermitian adjoint of the second arguments of the correlation functions in \cref{relation,phiijBouchard}. Since in the correlator \eqref{correlation} the second argument has a Hermitian adjoint, this ensures that the derivations of \cref{relation,phiijBouchard} are also applicable for the non-Hermitian spin ladder operators $S_+$ and $S_-$.

\nocite{apsrev41Control}
\bibliographystyle{apsrev4-1}
\bibliography{control,refs}

\end{document}